\documentclass[11pt]{article}
  
\usepackage{amsmath,amssymb}

\usepackage{epsfig}  
\usepackage{graphicx}               
\usepackage{url}
\usepackage{hyperref}

\newcommand{\nc}{\newcommand}  

\nc{\beq}{\begin{equation}}  
\nc{\eeq}{\end{equation}}  
\nc{\beqa}{\begin{eqnarray}}  
\nc{\eeqa}{\end{eqnarray}}  
\nc{\bea}{\begin{eqnarray}}  
\nc{\eea}{\end{eqnarray}}  
\nc{\ra}{\rightarrow}  
\nc{\lsim}{\begin{array}{c}\,\sim\vspace{-21pt}\\< \end{array}}  
\nc{\gsim}{\begin{array}{c}\sim\vspace{-21pt}\\> \end{array}}  
\nc{\Tr}{{\rm Tr}}
\nc{\slsh}{\slash\hspace*{-0.22cm}}
\def\eg{{\it e.g.}}

\def\be{\begin{equation}}
\def\ee{\end{equation}}
\def\bea{\begin{eqnarray}}
\def\eea{\end{eqnarray}}
\def\bit{\begin{itemize}}
\def\eit{\end{itemize}}

\newcommand{\missET}{\slash{\hspace{-2.5mm}E}_T}

\def\eg{{\it e.g.}}

\def\to{\rightarrow}

\title{  
\vspace*{-2.3cm}  
\begin{flushright}  
\normalsize{  
SLAC-PUB-14364 
  }  
\end{flushright}  
\vspace{1.5cm}  
\Large  
\textbf{
 LHC Predictions from a Tevatron Anomaly in the Top Quark Forward-Backward Asymmetry
 \\
}\vspace*{1.0cm}   
}

\author{Yang Bai, JoAnne L. Hewett, Jared Kaplan and Thomas G. Rizzo
\vspace{5mm}
\\
\normalsize\emph{SLAC National Accelerator Laboratory, 2575 Sand Hill Road, Menlo Park, CA 94025, USA}
}

\date{}

\begin{document}  
\setcounter{page}{0}  
\maketitle  

\vspace*{1cm}  
\begin{abstract} 
We examine the implications of the recent CDF measurement of the top-quark forward-backward asymmetry, focusing on a scenario with a new color octet vector boson at 1-3 TeV.  We study several models,  as well as a general effective field theory, and determine the parameter space which provides the best simultaneous fit to the CDF asymmetry, the Tevatron top pair production cross section, and the exclusion regions from LHC dijet resonance and contact interaction searches.  Flavor constraints on these models are more subtle and less severe than the literature indicates.  We find a large region of allowed parameter space at high axigluon mass and a smaller region at low mass; we match the latter to an $SU(3)_1\times SU(3)_2 / SU(3)_c$ coset model with a heavy vector-like fermion.  Our scenario produces discoverable effects at the LHC with only $1-2$ fb$^{-1}$ of luminosity at $\sqrt s = 7-8$ TeV.  Lastly, we point out that a Tevatron measurement of the $b$-quark forward-backward asymmetry would be very helpful in characterizing the physics underlying the top-quark asymmetry.
\end{abstract}  
\thispagestyle{empty}  
\newpage  
  
\setcounter{page}{1}

\baselineskip18pt   

\section{Introduction}
\label{sec:intro}

CDF recently announced a $3.4 \sigma$ deviation from the Standard Model (SM) in a measurement of the $t \bar t$ forward-backward asymmetry  \cite{Aaltonen:2011kc}.  Their study updates previous analyses that observed deviations of roughly $2 \sigma$ \cite{Aaltonen:2008hc,Abazov:2007qb}, but they also go further, demonstrating that the $t \bar t$ asymmetry is dominated by events with large invariant mass $M_{t \bar t}$.  This rise in the asymmetry at high energy could be interpreted as further evidence of a new physics signal characterized by energies at or above the weak scale.  In this work we will investigate explanations of the asymmetry based on new physics~\cite{Sehgal:1987wi}~\cite{Bagger:1987fz}\cite{Jung:2009jz, Ferrario:2009bz, Chivukula:2010fk, Cao:2010zb}, examine what can be done next at the Tevatron and the LHC to either rule out or confirm as well as characterize models that explain the anomaly.

An important first question is whether we should consider models that are wholly responsible for the asymmetry, or models where interference between new physics and QCD generates the asymmetry.  After unfolding their acceptance, CDF observed \cite{Aaltonen:2011kc} a parton-level asymmetry $A^{t \bar t}_{FB} =0.475\pm 0.114$ in $t \bar t$ events with $M_{t \bar t} > 450$ GeV and in the $t\bar t$ center of mass frame (see Table \ref{table:partondata}), compared to the SM prediction of $A^{t \bar t}_{FB}({\rm SM}) =0.088\pm 0.013$.    
This is a large effect.  If new physics alone generates the forward-backward asymmetry, a straightforward interpretation of these numbers suggests that new physics must be responsible for roughly half of all $t \bar t$ production above $450$ GeV, even in a scenario where the new physics yields a maximal forward-backward asymmetry!  However, a new heavy particle produced in the $s$-channel will have to contend with searches for resonances and contact interactions at both the Tevatron and the LHC \cite{Cao:2010zb}.  For this reason, we will focus on models where the new physics interferes with QCD to produce the asymmetry, allowing for a smaller total contribution to production rates from new physics.

In order to interfere with SM top production, new physics processes must produce $t \bar t$ pairs in a state with the same quantum numbers as in QCD production.  This means that the $t \bar t$ must form a color octet, Lorentz vector state, so if there is a new particle in the $s$-channel coupling to tops, it must be a heavy color octet vector boson\footnote{In principle, colored particles with spin $> 1$ might also interfere off-shell, but roughly speaking, this is due to the contribution of the vector particles they `eat' to become massive.}.  Such a `heavy gluon partner' $G'$ can arise in a wide variety of models \cite{Morrissey:2009tf}, including  top-color models~\cite{Hill:1991at}\cite{Hill:1993hs},  technicolor~\cite{Lane:1991qh}, warped extra dimensions~\cite{Randall:1999ee}\cite{Davoudiasl:1999tf}, universal extra dimensions~\cite{Appelquist:2000nn}  and chiral color~\cite{Pati:1974zv}\cite{Hall:1985wz}\cite{Frampton:1987dn}. As we will discuss below, $G'$ particles are constrained by flavor physics and collider searches, although we disagree with some exclusion claims in the literature.  For convenience we will present a phenomenological low-energy effective field theory and some simple coset models, and we will ignore the issue of anomaly cancellation. Throughout we will use the term `axigluon' as a shorthand for `color octet vector boson'.

New physics produced in the $t$-channel \cite{Jung:2009jz} could also generate a substantial forward-backward asymmetry.  Naively, one might imagine that a $t$-channel contribution would not rise with energy, but in fact $t$-channel physics produces a large energy dependence that might fit the CDF results nicely.  However, producing $t \bar t$ pairs via a new $t$-channel process generates non-trivial flavor issues; moreover, new $t$-channel effects already have significant tension with Tevatron data as they give rise to a large number of like-sign top pairs \cite{Jung:2009jz}.  Since the LHC is a $pp$ collider, a new $t$-channel state could be discovered soon by searches for like-sign tops.  We will not discuss these models any further here.

Future studies at both the Tevatron and the LHC could shed light on the nature of the $t \bar t$ asymmetry.  The Tevatron could provide a great deal of insight into the physics responsible for the $t \bar t$ asymmetry by studying a potential $b \bar b$ asymmetry, using semi-leptonic $b$ decays to tag the sign of $b$-jets.  If the heavy color octet $G'$ has large enough couplings to generate a sizable asymmetry, it should be visible in dijet resonance or contact interaction searches  
this year (with $1-2$ fb$^{-1}$ of luminosity) at the LHC.  In fact, we will see below that a sizable portion of the possible parameter space has already been excluded by the LHC!  Measurements of the $t \bar t$ production cross section are also sensitive to new physics, and a new $G'$ could be discovered as a resonance decaying to $t \bar t$.  Fixing the new physics contribution to $A^{t \bar t}_{FB}$, we will see that dijet and $t \bar t$ searches are complementary, as the product of the $G'$ coupling to light quarks and tops must essentially be held fixed to account for the asymmetry.  Whatever the verdict, in all likelihood new physics explanations for the asymmetry will be much better understood after the next year of operations at the LHC.

\begin{table}
\renewcommand{\arraystretch}{1.5}
\center
\begin{tabular}{| c | c | c |} 
\hline  
Selection & $M_{t \bar t} < 450$ GeV  &  $M_{t \bar t} > 450$ GeV \\ 
\hline  
Parton Level Data & $-0.116 \pm 0.146 \pm 0.047$ & $0.475 \pm 0.101 \pm 0.049$  \\ 
\hline 
\hline 
Selection & $| \Delta y | < 1.0$ & $| \Delta y | > 1.0$ \\ 
\hline  
Parton Level Data & $0.026 \pm 0.104 \pm 0.056$ & $0.611 \pm 0.210 \pm 0.147$ \\ 
\hline
\end{tabular}
\caption{Unfolded parton level data for $A^{t \bar t}_{FB}$ from the CDF study \cite{Aaltonen:2011kc}.}
\label{table:partondata}
\end{table}

The rest of this paper is organized as follows.  Section 2 introduces an effective field theory for a heavy color octet vector boson and then discusses various collider bounds that constrain it.  In Section 3 we introduce specific coset models for $G'$, and define the parameter space for each model that provides a fit to the CDF asymmetry data at 90\% and 95\% C.L.  We also show how the constraints from other collider data ($\sigma_{t\bar t} $, dijet resonance, and dijet contact interaction searches) impact this parameter space.  In Section 4 we explore the contributions in the flavor sector from our models and show that meson mixing does not necessarily restrict the relevant parameter space.  Section 5 delineates various signatures of the heavy color octet at the LHC.  Lastly, Section 6 contains our conclusions.

\section{An Axigluon?}
\label{sec:constraints}

Our goal is to examine if a new color octet vector boson $G'$ can explain the discrepancy in the recently measured $t \bar t$ forward-backward asymmetry.  To this end, let us begin with a general low-energy effective Lagrangian coupling $G'$ to the SM quarks.  To suppress dangerous FCNC effects it will prove advantageous to assume universal vector and axial couplings to the first two generations, $g_V^q$ and $g_A^q$, while the couplings to the top quark and bottom quark will be $g_V^t$ and $g_A^t$.  For convenience we have rescaled these couplings in terms of the strong coupling $g_s$.  The tree-level differential cross section for $q \bar q \to t \bar t$ will be \cite{Hewett:1988xc,Ferrario:2009bz} 
\bea
\label{eqn:phenom}
\frac{d \hat{\sigma}^{q \bar q \to t \bar t}}{d \cos \theta^*} &=& \alpha_s^2 \frac{\pi \sqrt{1-4m^2}}{9 \hat{s}} \left[ \left(1 + 4m^2 + c^2\right) \left(1 -  \frac{2  g_V^q g_V^t  \hat{s}(M_{G^\prime}^2 - \hat{s}) }{(\hat{s}-M_{G^\prime}^2)^2+M_{G^\prime}^2 \Gamma_G^2}   
+  \frac{g_V^{t\,2}  (g_V^{q\,2} + g_A^{q\,2}) \hat{s}^2}{(\hat{s}-M_{G^\prime}^2)^2+M_{G^\prime}^2 \Gamma_G^2}  \right)   \right. \nonumber \\ 
& & + \left(1 - 4m^2 + c^2 \right) g_A^{t\,2}  (g_V^{q\,2} + g_A^{q\,2}) \frac{\hat{s}^2}{(\hat{s}-M_{G^\prime}^2)^2+M_{G^\prime}^2 \Gamma_G^2} \nonumber \\
& & - \left.  4 g_A^q g_A^t c \left(  \frac{\hat{s}(M_{G^\prime}^2-\hat{s}) }{(\hat{s}-M_{G^\prime}^2)^2+M_{G^\prime}^2 \Gamma_G^2}  -  2 g_V^q g_V^t \frac{\hat{s}^2}{(\hat{s}-M_{G^\prime}^2)^2+M_{G^\prime}^2 \Gamma_G^2} \right) \right]
\eea
where $m^2 = m_t^2/\hat{s}$, $c = \sqrt{1-4m^2} \cos \theta^* $, and $\theta^*$ is the angle between the top quark and  the incoming quark in the center of mass frame.  The forward-backward asymmetry arises solely from the last line in this equation, so to obtain a positive asymmetry we must have $g_A^q g_A^t < 0$.

We will be considering $M_{G^\prime} \gtrsim 1$ TeV, so at the Tevatron we have $\hat s < M_{G^\prime}^2$ over most of the support of the parton distribution functions.  In this limit the interference term between the QCD and resonance contributions dominates over the pure resonance term, and we expect the following qualitative features:
\begin{itemize}
\item the asymmetry is directly proportional to $(-g_A^q g_A^t)$
\item relative to the pure QCD cross section, the asymmetric term grows $\propto \hat{s}$, so the asymmetry will grow with the invariant mass squared $M_{t \bar t}^2$  
\item the total $t \bar t$ cross section at low energies will decrease with increasing $g_V^q g_V^t$.
\end{itemize}
Additionally, we see that the sign and magnitude of $g_V^q g_V^t$ determine the behavior of the asymmetry at high energies, and that for larger $g_V^q g_V^t$, our qualitative discussion based on the parametric expansion in $\hat s/M_{G^\prime}^2$ breaks down more quickly.  

Axigluons face constraints from flavor-changing neutral currents (FCNCs), from $t \bar t$ resonance searches, and from measurements of the total $t \bar t$ cross section.  Furthermore, any axigluon that couples to $q \bar q$ is subject to constraints from dijet resonance and dijet contact interaction searches -- not just at the Tevatron but at the LHC!  As we will see, the strongest constraints on our models will come from the LHC, and the prospects of a discovery or a definitive exclusion this year are extremely good.  This is striking proof that we have entered the LHC era!

Before considering the details, note that at least qualitatively, once we fix $g_A^q g_A^t$, the $t \bar t$ and dijet constraints will be orthogonal.  Increasing $|g_A^q g_A^t|$ increases the asymmetry, but eventually one runs afoul of constraints; one might avoid altering the $t \bar t$ cross section by increasing $g_V^q g_V^t$, but only at the expense of decreasing the asymmetry itself.  Now let us discuss the restrictions from the $t \bar t$ cross section and the dijet searches; we will leave a discussion of flavor to Section \ref{sec:flavor}.

\subsection{The $t \bar t$ Cross Section}
\label{sec:ttbarcrosssection}

Theoretical predictions \cite{Cacciari:2008zb} of the $t \bar t$ cross section have been verified at the Tevatron \cite{CDFttbarProduction}.  An axigluon will modify both the total $t \bar t$ cross section and its shape as a function of the invariant mass $M_{t \bar t}$, so Tevatron $t \bar t$ measurements restrict axigluon models.  However, we cannot set a precise constraint without a more detailed understanding of the experimental and theoretical details.  The current $t\bar{t}$ resonance searches at the Tevatron~\cite{Abazov:2008ny}\cite{D0ttbar}\cite{CDFttbar}\cite{Aaltonen:2009iz}\cite{Aaltonen:2009tx} have not set a limit on a possible resonance with mass beyond $\sim 800$~GeV due to a lack of statistics in their published analyses.

In our analysis below, we will display contours corresponding to an axigluon contribution  of $20 \%$ and $40 \%$ to the total QCD $t \bar t$ cross section for  $M_{t\bar{t}} > 450$~GeV.  These contours should be viewed as a suggestion of the Tevatron sensitivity, and not as hard bounds.  One might instead consider using a larger cut on the top invariant mass.  This would result in a larger relative axigluon contribution to the cross section, but would compete with the uncertainties on the QCD cross section which grow rapidly at higher invariant mass.  After some experimentation, we chose this $M_{t\bar{t}} > 450$~GeV requirement as a reasonable and conservative compromise between these two factors.

For illustrative purposes, we show the $M_{t\bar{t}}$ distributions in the upper panel of Fig.~\ref{fig:dijetbroad}. With the exception of the magenta line, the other curves on the histogram represent a 1 TeV axigluon. For the red(blue) lines, the coupling of the axigluon is of QCD strength and $\Gamma/M = 0.1(0.2)$ has been assumed.  For the green(cyan) histogram, the coupling strengths of the axigluon have been doubled and now $\Gamma/M = 0.4(0.8)$ has been assumed. One can see that a 1 TeV axigluon with a narrow width will appear as a $t\bar{t}$ resonance.  

\subsection{Dijet Resonance Searches}
\label{sec:dijetresonance}

CMS recently performed a search for narrow dijet resonances \cite{Khachatryan:2010jd}, employing an integrated luminosity of 2.9~pb$^{-1}$.  This search has ruled out colorons and axigluons (with couplings $g_V^{q, t} = 1$ or $g_A^{q, t} = 1$ respectively, in our notation) in the mass intervals $0.50 < M_{G^\prime} < 1.17$~TeV and $1.47 < M_{G^\prime} < 1.52$~TeV.  In the narrow width approximation the production cross section 
for $q\bar{q} \rightarrow G^\prime \rightarrow q\bar{q}$  is 
\be
\frac{\sigma \times BR(q \bar q \to G' \to jj)}{\sigma \times BR(\mathrm{coloron} ) } 
=  \frac{6}{5} \left(|g_V^q|^2 + |g_A^q|^2 \right)  \frac{4(|g_V^q|^2 + |g_A^q|^2) + (|g_V^t|^2 + |g_A^t|^2)}{4(|g_V^q|^2 + |g_A^q|^2)+2(|g_V^t|^2 + |g_A^t|^2)}  \,,
\ee  
where we have related it to the cross section for a coloron with coupling $g_s$, and we have assumed an identical coupling to bottom and top quarks.  Using the CMS 95\% C.L. exclusion upper limit on dijet resonance production, $\sigma\times A$ (where $A$ is the acceptance), for the $q \bar q$ initial state and then rescaling the cross sections of colorons and axigluons in Figure 4 of Ref.~\cite{Khachatryan:2010jd}, we can estimate the corresponding dijet resonance constraints~\cite{Han:2010rf}. 

\begin{figure}[th!]
\begin{center}
\vspace*{-2.9mm}
\includegraphics[width=0.6\textwidth]{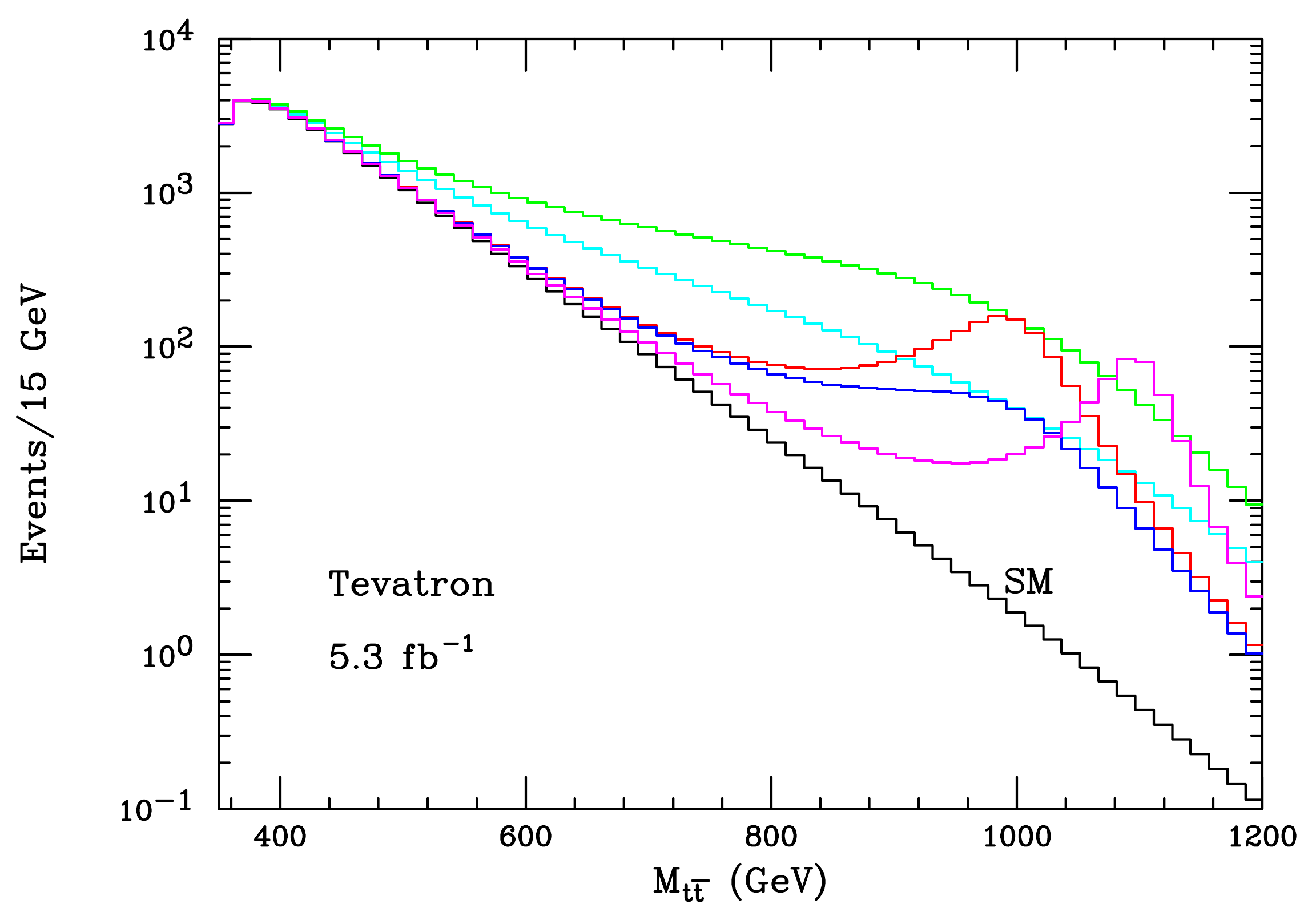} 
\\
\vspace{2mm}
\includegraphics[width=0.6\textwidth]{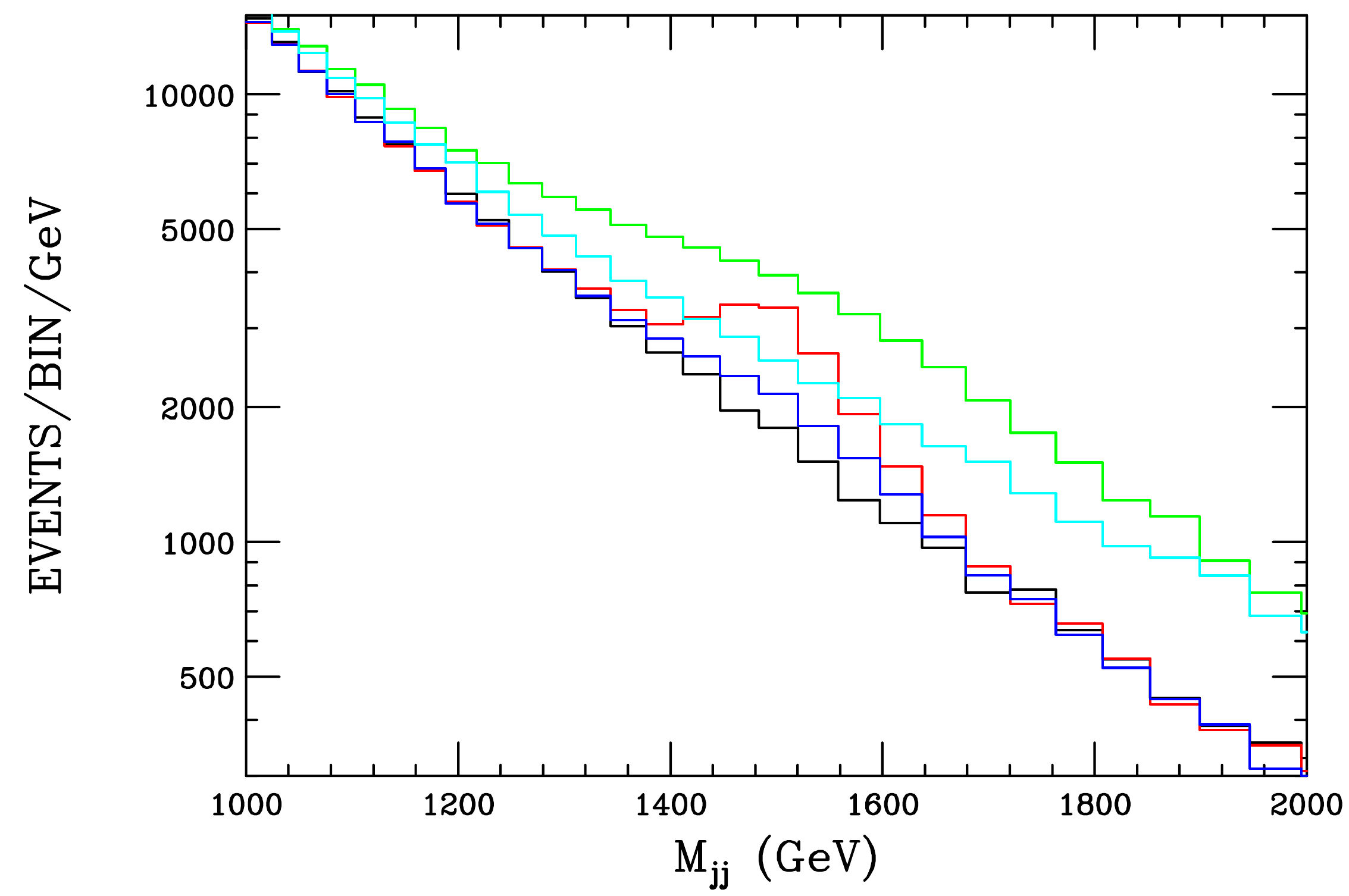} 
\vspace{0cm}
\caption{Upper panel: the axigluon contribution to the $t\bar{t}$ invariant mass distribution at the Tevatron. No $t$-quark branching fractions or reconstruction efficiencies have been employed. The magenta curve represents a model we will discuss later in Section~\ref{sec:vectorfermion} with parameter values $\theta=30^\circ$ and $M_{G^\prime}$ =1100 GeV.
The black histogram is the QCD prediction whereas the red(blue) histogram assumes the 
existence of a 1.0 TeV axigluon with QCD strength couplings with 
$\Gamma/M \simeq 0.1(0.2)$. For the green(cyan) histogram, the coupling 
strengths of the axigluon have been doubled and now $\Gamma/M \simeq 0.4(0.8)$ has been assumed. The CTEQ6.6M PDFs were employed. 
Lower panel:  axigluon contribution to the dijet pair production cross section at the LHC 
assuming $\sqrt s=8$ TeV and  an integrated luminosity of L=5~fb$^{-1}$ 
after applying the cuts $|\eta_j|<0.5$, $p_{Tj}>200$ GeV and CMS detector smearing.
All other four  colored histograms are  for a 1.5 TeV axigluon with the same $\Gamma/M$ ratios as the upper panel.}
\label{fig:dijetbroad}
\end{center}
\end{figure}

The dijet resonance searches are based on the narrow width approximation, so they do not constrain axigluons with large widths. 
To give an idea of when the resonance has become too broad for this search, in the lower panel of 
Fig.~\ref{fig:dijetbroad} we show a sample of  LHC dijet invariant mass distributions for the signal and QCD background.  As one can see from this figure, for $\Gamma/M \simeq 0.1$ (the red line) the signal shows a bump-like feature on top of the QCD background (the black line). However, for $\Gamma/M \gtrsim 0.2$ (the blue line) the bump is much less visible. We will thus assume in what follows that the narrow resonance search becomes inapplicable when $\Gamma/M \gtrsim 0.2$.  

\subsection{Dijet Contact Interaction Searches}
\label{sec:dijetcontact}

Searches for quark contact interactions can be sensitive to broad axigluons via dijet angular distributions. This follows because LO QCD dijet 
production is dominated by $t$-channel Rutherford-like scattering. For a small production angle $\theta^*$ in the center of mass frame, the QCD 
differential cross section behaves as $d\hat{\sigma}/d \cos{\theta^*} \sim 1/\sin^4(\theta^*/2)$. Defining the variable 
$\chi = (1+\cos{\theta^*})/(1-\cos{\theta^*})$ and considering the limit $\chi \rightarrow \infty$, one finds that  $d\hat{\sigma}/d\chi \sim$ constant. 
On the other hand, the differential distribution for dijets arising from resonance production is not flat in $\chi$ but instead has a peak at small values of $\chi$. 
By comparing the corresponding shapes of the signal and background distributions in $\chi$, ATLAS has excluded LL quark contact interactions\footnote{The use of a left-left contact interaction is a common convention or benchmark; the limits on a L-R or R-R interaction would be similar in magnitude.} with a 
compositeness scale below 3.4~TeV at 95\% C.L.~\cite{Collaboration:2010eza}.  To set corresponding limits on our model parameter space, we require 
that the new physics contribution plus the SM background not exceed the ATLAS data at  95\% C.L. for their chosen mass intervals $800< m_{jj} < 1200$~GeV and  
$m_{jj} > 1200$~GeV. Only the first four ATLAS bins with $1< \chi  < 3.32$ are included in this analysis, which we base on 
the optimized sensitivity to quark contact interactions given in Ref.~\cite{Collaboration:2010eza}.  As an example of this search in action, the constraints from dijet contact interaction searches 
are shown as the solid dark yellow line of Fig.~\ref{fig:generalmodelratio}, which refers to the effective field theory of section \ref{sec:constraints}. 

The reader will notice that the contact interaction  search always provides a tighter constraint on our models than the dijet resonance search.  This follows because the narrow resonance analyses only examine the shape difference without subtracting QCD background, while the contact interaction  search has subtracted the QCD background and hence should constrain new physics more. Our constraints from dijet contact interactions are conservative because we neglected the NLO QCD correction. To have a more precise constraint, one should perform a NLO calculation as done in Ref.~\cite{Gao:2011ha}.

\subsection{Phenomenological Fit}

We first perform a model independent fit to $A^{t\bar{t}}_{FB}$ with one axigluon field, with a cross section given by Eq.~(\ref{eqn:phenom}). In this case, there are five parameters that describe the axigluon contributions to $A^{t\bar{t}}_{FB}$ and the $t\bar t$ production cross section. Instead of scanning all of this parameter space, we set $g_V^q = g_V^t =0$ in this section and only consider three parameters: $M_{G^\prime}$, $g_A^q$ and $g_A^t$, which are most crucial for $A^{t\bar{t}}_{FB}$ when $M_{t\bar t} \ll M_{G^\prime}$. As discussed above, a negative value for the product $g_A^qg_A^t$ provides a positive contribution to $A^{t\bar{t}}_{FB}$. Therefore, we restrict the parameter space to  $g_A^q > 0$ and $g_A^t < 0$ (the opposite choice would not change our discussion below).

\begin{figure}[t]
\begin{center}
\vspace*{3mm}
\includegraphics[width=0.45\textwidth]{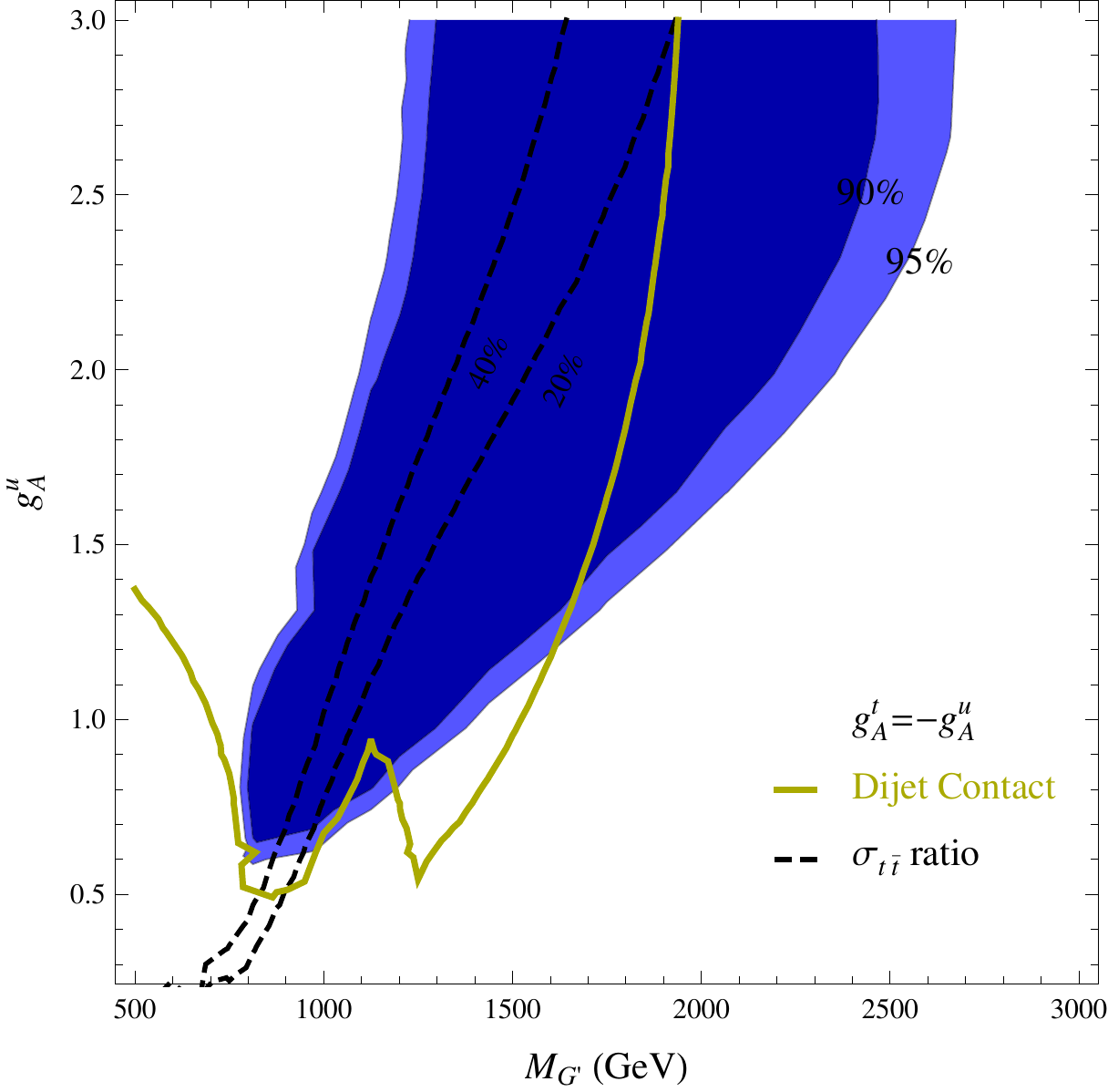} \hspace{1cm}
\includegraphics[width=0.45\textwidth]{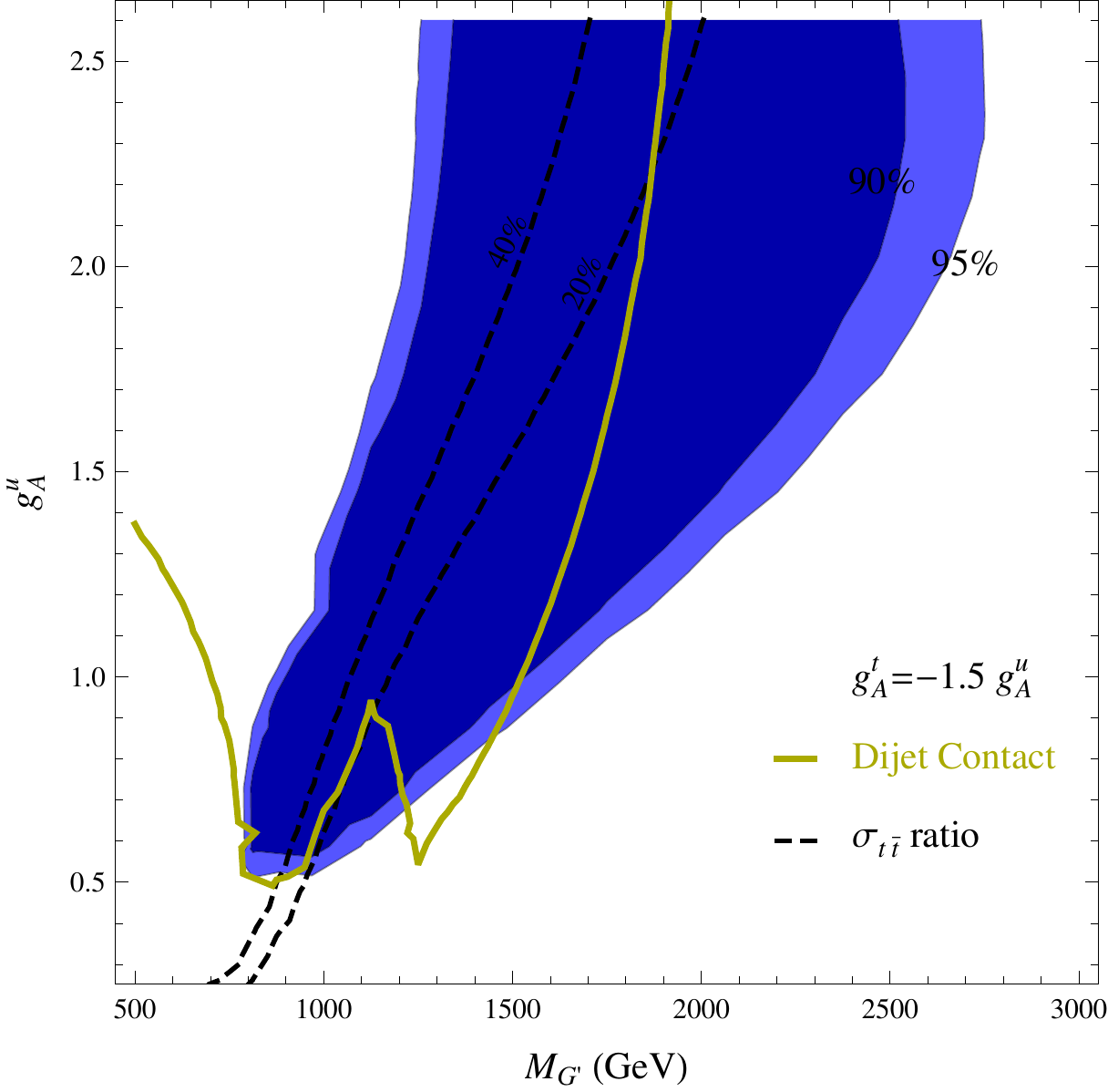} 
\caption{Left panel: the fit for the phenomenological model of Eq.~\ref{eqn:phenom} with $g_A^q > 0$, $g_A^t < 0$, $g_V^{q, t}=0$ and a fixed ratio $g_A^t /g_A^q = -1$. The region above and to the left of the dark yellow solid line is excluded by dijet contact  interaction searches; limits from the dijet resonance searches are weaker.  Right panel: the same as on the left but now for $g_A^t /g_A^q = -1.5$.  The two blue shaded regions correspond to a 90\% and 95\%
CL fit to the data.}
\label{fig:generalmodelratio}
\end{center}
\end{figure}

\begin{figure}[t]
\begin{center}
\vspace*{3mm}
\includegraphics[width=0.45\textwidth]{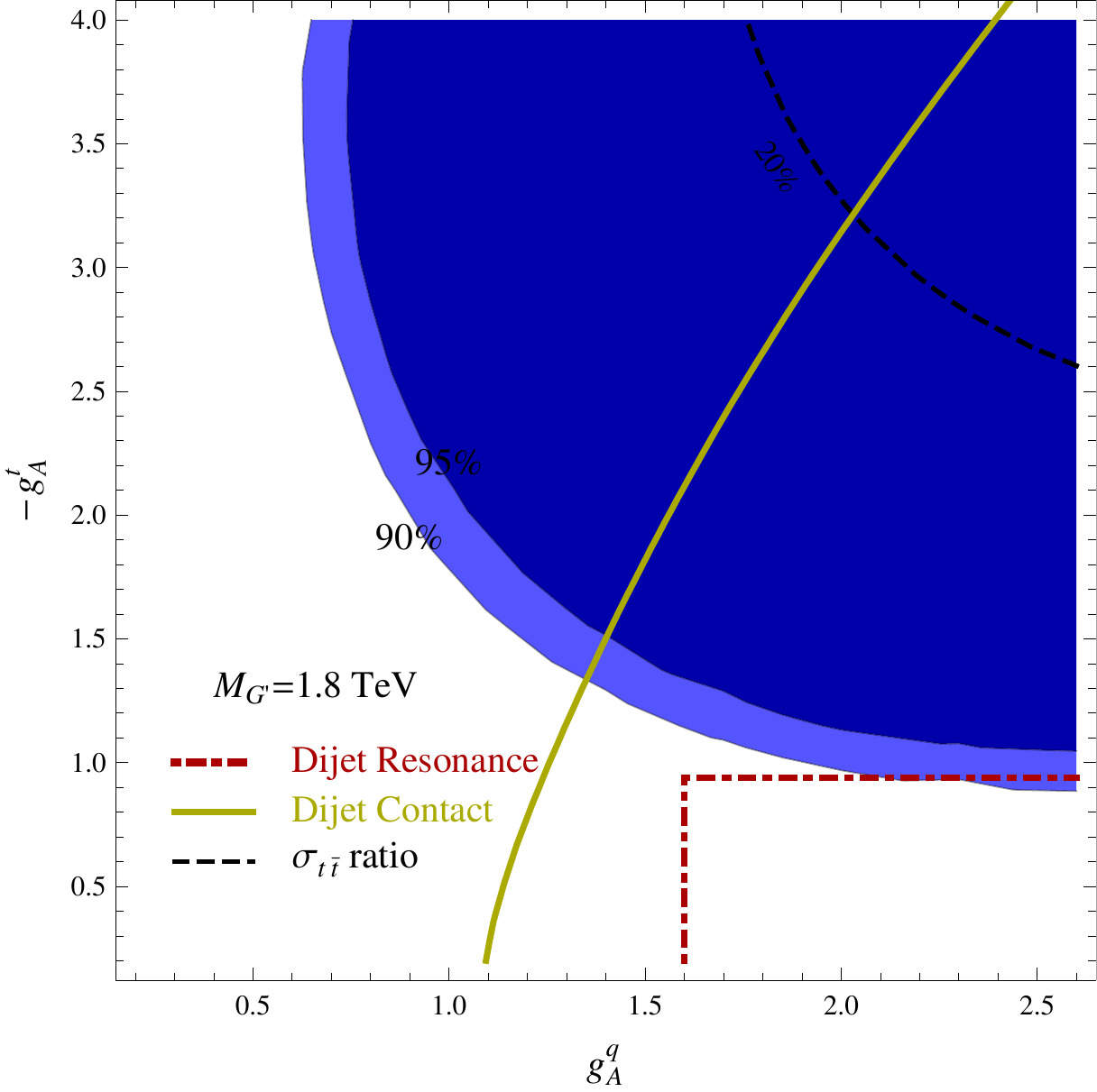} \hspace{1cm}
\includegraphics[width=0.45\textwidth]{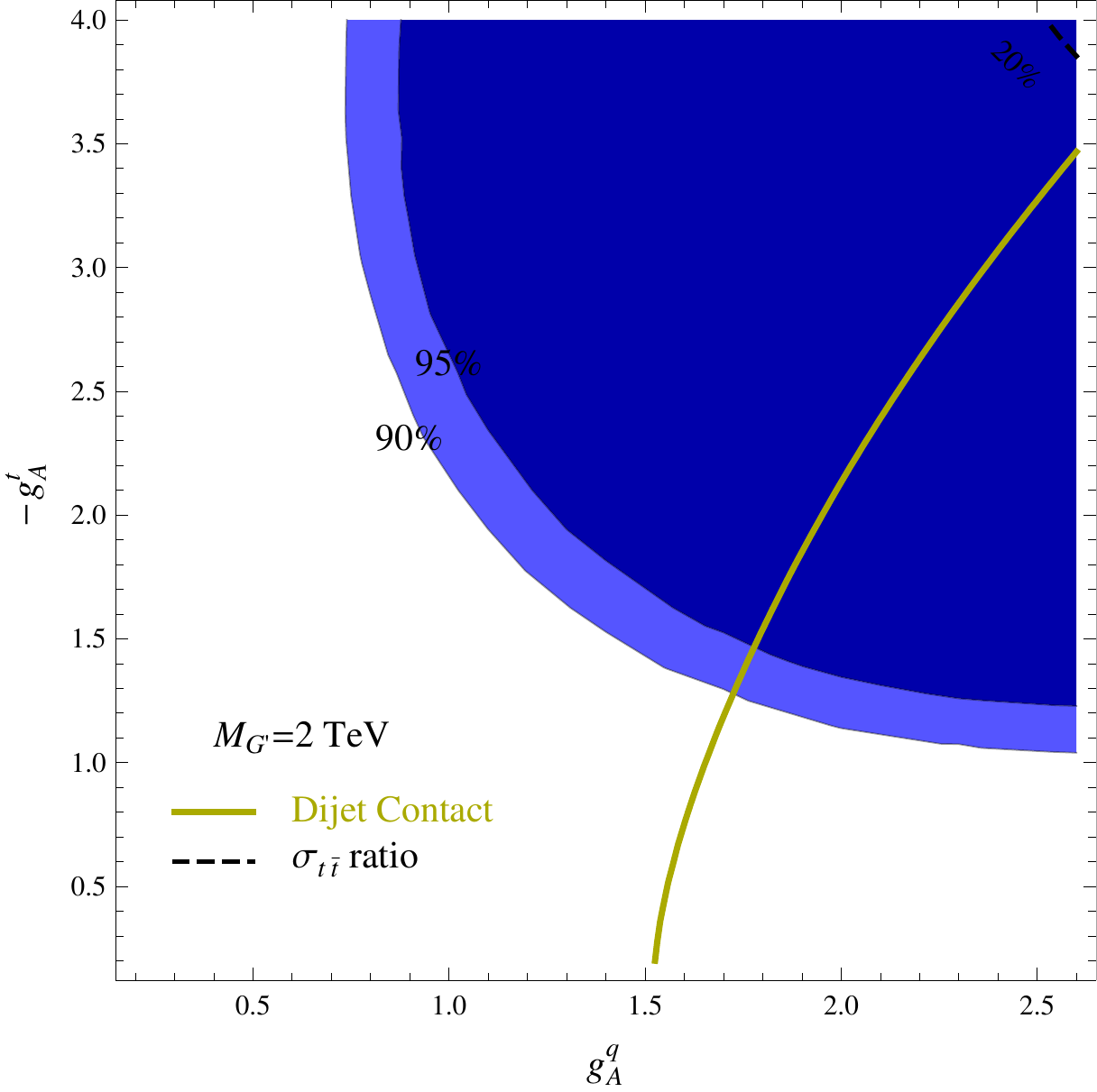} 
\caption{Left panel: the fit for the phenomenological model  of Eq.~\ref{eqn:phenom} with $g_A^q > 0$, $g_A^t < 0$ and $g_V^{q, t}=0$. The axigluon is assumed to have a mass of 
$M_{G^\prime} =1.8$~TeV. The region on the right side of the dark yellow solid line is excluded by dijet contact interaction searches.  Right panel: the 
same as on the left but now for $M_{G^\prime} =2$~TeV. The narrow resonance dijet searches do not further constrain the parameter space.}
\label{fig:generalmodel}
\end{center}
\end{figure}

In the contours displayed in Fig.~\ref{fig:generalmodelratio} we fix the ratio $g_A^t/g_A^q$. In the left panel, we choose $g_A^t = -g_A^q$ and show the 90\% and 95\% C.L. fit with two shaded blue regions. The dark yellow  solid line represents the constraint from dijet contact interaction  searches. All parameter regions above and to the left of this line are excluded. The black dashed lines correspond to possible $20\%$ and $40\%$ axigluon contributions to the top quark pair production cross section, relative to that of the LO SM, for $M_{t\bar{t}} > 450$~GeV, as discussed in Section \ref{sec:ttbarcrosssection}. The upper limit for $g_A^q$ in this figure is chosen so that the $G^\prime$ width does not exceed its mass. The right panel in this figure is similar to the left but now for $g_A^t = -1.5 g_A^q$. As one can see from this figure, the low mass region is mostly ruled out by the dijet contact interaction  searches, but a higher mass $G^\prime$ has plenty of parameter space. Concentrating on the high mass region, we next fix the $G^\prime$ mass and show the fit contour regions in Fig.~\ref{fig:generalmodel}, where the left(right) panel is for $M_{G^\prime} = 1.8(2.0)$ TeV.
Comparing those two panels, one can see that more parameter space survives for a heavier $G^\prime$, which also has larger couplings and hence a larger 
width in the best-fit region. As an example, we show the fit results for $A^{t\bar t}_{FB}$ in Table~\ref{table:fitphenomodel} with $M_{G^\prime} = 2$~TeV, $g_A^q = 2.2$, $g_A^t = - 3.2$ and $g_V^{q, t}=0$.  
\begin{table}
\renewcommand{\arraystretch}{1.5}
\center
\begin{tabular}{| c | c | c |} 
\hline  
Selection & $M_{t \bar t} < 450$ GeV  &  $M_{t \bar t} > 450$ GeV \\ 
\hline  
Parton Level Exp. Data & $-0.116 \pm 0.146 \pm 0.047$ & $0.475 \pm 0.101 \pm 0.049$  \\ 
\hline 
Model Prediction & $0.10$ & $0.31$  \\ 
\hline 
\hline 
Selection & $| \Delta y | < 1.0$ & $| \Delta y | > 1.0$ \\ 
\hline  
Parton Level Exp. Data & $0.026 \pm 0.104 \pm 0.056$ & $0.611 \pm 0.210 \pm 0.147$ \\ 
\hline 
Model Prediction & $0.12$ & $0.40$  \\ 
\hline
\end{tabular}
\caption{The comparison of theoretic predictions and measured values for the phenomenological model with $M_{G^\prime} = 2$~TeV, $g_A^q = 2.2$, $g_A^t = - 3.2$ and $g_V=0$. The total $\chi^2$ is 5.5.}
\label{table:fitphenomodel}
\end{table}

\section{Axigluon Models}

In this section we will discuss three specific axigluon models that can be described in the language of dimensional deconstruction, or simply as coset models.  The first and third are based on cosets $[SU(3) \times SU(3)] / SU(3)_c$, while the second involves three $SU(3)$ groups and two new color octet vector bosons.  The third model differs from the first by the introduction of new vector-like quarks.

These models have elementary motivations.  The first is the simplest possible coset model that produces an axigluon mediated $t \bar t$ forward-backward asymmetry, but we find it is excluded by the ATLAS dijet contact interaction  search \cite{Collaboration:2010eza}.  The second model improves on the first by allowing for heavier and more strongly coupled axigluons, yet we will see that it is also excluded.  The third model makes it possible to avoid the powerful dijet contact interaction search \cite{Collaboration:2010eza} by reducing the axigluon coupling to light quarks while increasing the axigluon coupling to $t \bar t$, thereby keeping the asymmetry fixed.  Since constraints on $t \bar t$ production are weaker and less precise than the constraints on dijets, our third model survives -- although it will be tested at the LHC in the coming months!

The models that follow provide calculable perturbative examples, but a more general theory such as our effective Lagrangian in Section \ref{sec:constraints} might be the best new physics explanation for the $t \bar t$ asymmetry, in which case the asymmetry may be the first hint of new strongly coupled physics.

\subsection{The Minimal Axigluon Model}
\label{sec:twosites}
The simplest way to introduce another color octet vector boson is to extend the QCD gauge symmetry to the group $SU(3)_1 \times SU(3)_2$, which then 
spontaneously breaks to its subgroup $SU(3)_{c}$. This spontaneous symmetry breaking can be achieved by introducing a complex scalar field 
$\Sigma$ which transforms as a $(3, \bar{3})$ under the $SU(3)_1 \times SU(3)_2$ symmetry with a non-zero VEV, 
$\langle\Sigma\rangle =\frac{\mathbb{I}_3}{\sqrt{6}}\,f_\Sigma$.  There are only two parameters 
in this model and they can be identified as the axigluon mass $M_{G^\prime}$ and the gauge coupling mixing angle $\theta$. Here, we only consider 
$0\leq \theta < 45^\circ$, since the range $90^\circ - \theta$ can be mapped into the region $0\leq \theta < 45^\circ$ by flipping the quark charges 
under $SU(3)_1$ and $SU(3)_2$. 
Diagonalizing the mass matrix of the two gauge bosons $G_1^\mu$ and $G_2^\mu$, we obtain the massless QCD gluon,
\beq
G^\mu = \cos{\theta}\,G_1^\mu +  \sin{\theta}\,G_2^\mu \,,
\eeq
and the massive axigluon state,
\beq
G^{\prime\,\mu} = - \sin{\theta}\,G_1^\mu + \cos{\theta}\,G_2^\mu \,.
\eeq
The mixing angle $\theta$ is related to the gauge couplings of $SU(3)_1 \times SU(3)_2$, $h_1$ and $h_2$: $\tan{\theta} = h_1/h_2$. The QCD 
coupling is then given by $g_s = h_1\cos{\theta}= h_2\sin{\theta}$, and the mass of the axigluon is 
\beq
M_{G^\prime} = \frac{\sqrt{2}\,g_s}{\sqrt{3} \sin{2\theta}}\,f_\Sigma \,.
\eeq
The other degrees of freedom in $\Sigma$ are assumed to be heavy for now, so we only have one new particle $G^\prime_\mu$ below the scale  
$\sim 4\pi f_\Sigma$.

In order to obtain an $A^{t\bar{t}}_{FB}$ with the correct sign, we need opposite signs for the axial-vector couplings of the axigluon to the 
light quarks and the top quark. To achieve this goal, we make the following assignments for the SM quarks under $SU(3)_1\times SU(3)_2$: $q_L$, 
$t_R$, $b_R$ as triplets of $SU(3)_1$ and $(t, b)_L$, $q_R$ as triplets of $SU(3)_2$ (see Ref.~\cite{Frampton:2009rk} for a similar setup). To cancel the gauge anomalies, additional colored particles are required and are assumed to be heavy in here.
Here, ``$q$" represents the first two generations of quarks.  With these charge assignments, we find the vector and  axial-vector couplings of 
$G^\prime_\mu$ to the SM quarks, re-scaled by the QCD coupling $g_s$,  to be
\beq
g^t_V = g^q_V =  \frac{1}{\tan{2\theta}} \,, \qquad 
g^t_A = - g^q_A =  \frac{1}{\sin{2\theta}} \,.
\label{eq:twositecoupling}
\eeq
Neglecting the quark masses (since $M_{G^\prime}$ will be at or above the TeV scale), the total decay width is found to be
\beq
\Gamma(G^\prime) = \frac{\alpha_s\,M_{G^\prime}}{6}\left[ 4 (|g^q_V|^2 + |g^q_A|^2 ) + 2(|g^t_V|^2 + |g^t_A|^2) \right]  = \alpha_s\,M_{G^\prime} \, 
\left( \frac{1}{\tan^2{2\theta}}  + \frac{1}{\sin^2{2\theta}}  \right)\,.
\label{eq:width}
\eeq
For $\theta = \pi/4$, one finds that the width to mass ratio of $G^\prime$ is $\Gamma(G^\prime)/M_{G^\prime} = \alpha_s \sim10\%$. We also note 
that if $G^\prime$ has other decay channels, it could become a broader resonance, a possibility we will consider later. 

\begin{figure}[t]
\begin{center}
\vspace*{3mm}
\includegraphics[width=0.6\textwidth]{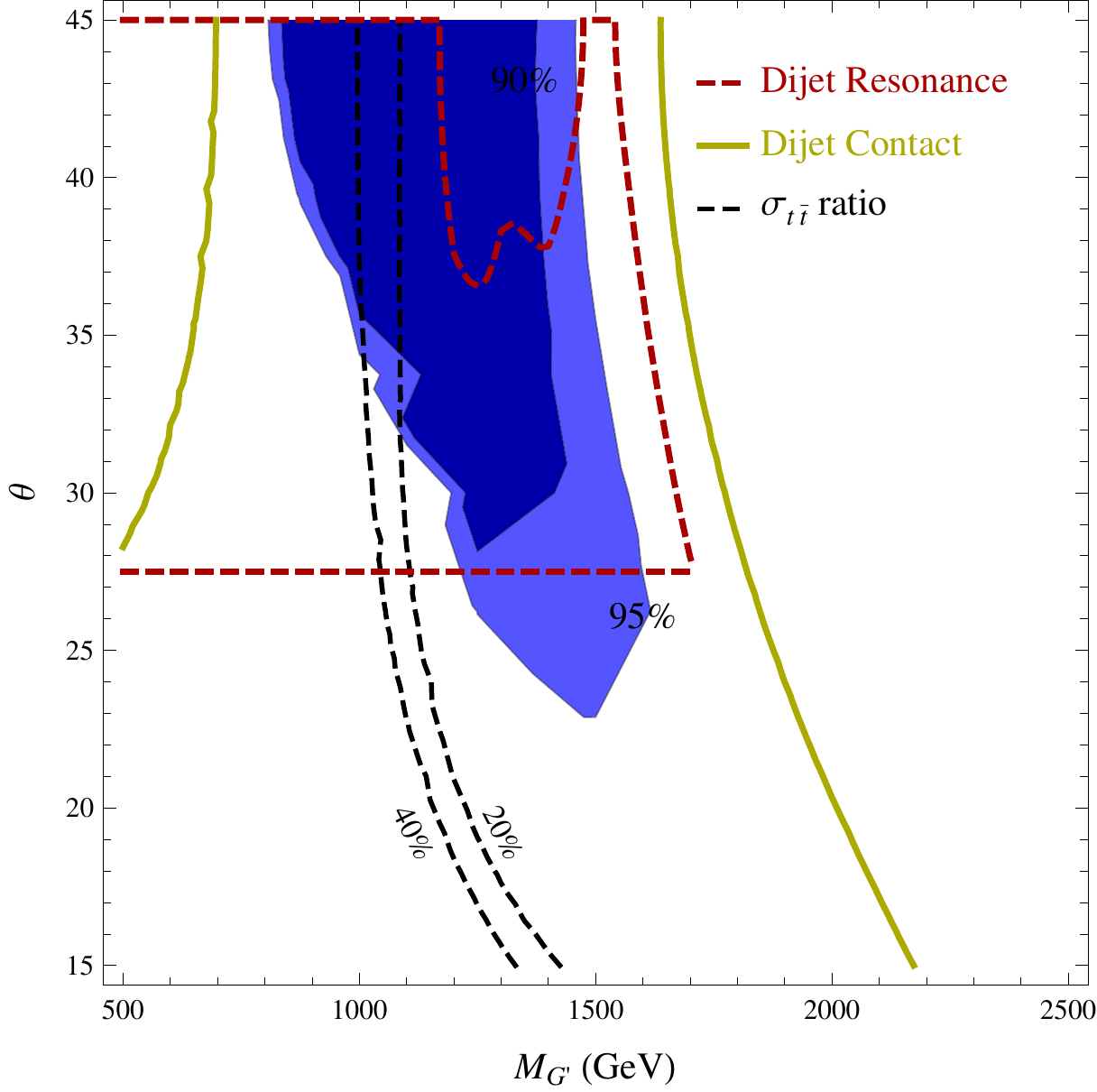} 
\caption{The fit of the minimal two-site axigluon model to the observed $A^{t\bar{t}}_{FB}$ at CDF. The four-bin values of $A^{t\bar{t}}_{FB}$ corresponding to 
high and low rapidity and high and low invariant mass are included in the $\chi^2$ analysis. The best fit point has a $\chi^2/d.o.f.= 5.3/2$ 
occurring at $M_{G^\prime} = 1041$~GeV and $\theta =45^\circ$. The region enclosed by the red dashed lines is excluded by the dijet narrow resonance search 
at CMS~\cite{Khachatryan:2010jd} with 2.9 pb$^{-1}$ luminosity. The region between the dark solid yellow lines is excluded by the search for diquark 
contact  interactions from ATLAS~\cite{Collaboration:2010eza} with 3.1 pb$^{-1}$, so all of the preferred parameter space of this model has been eliminated.}
\label{fig:twosite}
\end{center}
\end{figure}

The interference term in $q\bar q\to t\bar{t}$ production between the $s$-channel gluon and $G^\prime$ contains a term linear in the cosine of the production 
angle of the $t$ in the $t\bar{t}$ rest frame; this leads to a forward-backward asymmetry (see Eq.~(\ref{eqn:phenom}) and Ref.~\cite{Ferrario:2009bz}). 
We perform a fit to the observed and unfolded parton-level four-bin data from Ref.~\cite{Aaltonen:2011kc}, with the results being shown in Fig.~\ref{fig:twosite}, where 
the two blue contour regions correspond to the 90\% and 95\% confidence level, respectively. In our fit, we have neglected the NLO SM contributions to  
$A^{t\bar{t}}_{FB}$ since we 
only include the leading order contribution from the new model; note that the radiative NLO SM asymmetry aligns with the asymmetry from the axigluon. 
The best-fit point is found to be at $M_{G^\prime}=1041$~GeV and $\theta = 45^\circ$ with 
$\chi^2/d.o.f = 5.3/2$ for four bins, compared to $\chi^2= 24.2/4$ for the SM tree-level only fit.
The black dashed lines in  Fig.~\ref{fig:twosite}, correspond to possible $20\%$ and $40\%$ axigluon contributions to the top quark pair production 
cross section, relative to that of the LO SM, for $M_{t\bar{t}} > 450$~GeV, as we discussed in Section \ref{sec:ttbarcrosssection}. 

The axigluon couples to quark pairs with a coupling as strong as in QCD, therefore searches for possible dijet resonances \cite{Khachatryan:2010jd} will further 
constrain the parameter space of the model, as discussed in Section \ref{sec:dijetresonance}.  We note that in the narrow width approximation the production cross section 
for $q\bar{q} \rightarrow G^\prime \rightarrow q\bar{q}$ is proportional to $|g_V^q|^2 + |g_A^q|^2 = \cot^2{2\theta} + \csc^2{2\theta}$ and that the 
branching fraction for $G^\prime$ decaying into jets is fixed to be 5/6.  Hence, as $\theta$ moves away from $45^\circ$, the production cross section increases and a stringent constraint can be obtained.  Employing the CMS 95\% C.L. exclusion upper limit on the dijet resonance production $\sigma\times A$ from $q\bar q$ initial processes and then rescaling the cross sections for colorons and axigluons from 
Figure 4 of Ref.~\cite{Khachatryan:2010jd}, we present the corresponding narrow dijet resonance constraints as the red dashed line in Fig.~\ref{fig:twosite}. 
Since these dijet resonance searches are based on the narrow width approximation, they will not constrain axigluons with large widths, as we saw with Fig.~\ref{fig:dijetbroad}.

Although a broad $G^\prime$ may survive the dijet resonance analyses, searches for quark contact interactions \cite{Collaboration:2010eza} via the dijet 
angular distributions can be used to constrain our model parameter space even when the $G^\prime$ is broad. The constraints from the dijet contact interaction  
searches are shown as the solid dark yellow lines of Fig.~\ref{fig:twosite}. The parameter space enclosed by these two lines is excluded 
by this search. One can see that all of the $A^{t\bar{t}}_{FB}$ preferred parameter space regions are now ruled out.  Therefore, we conclude that the minimal axigluon model can not adequately 
explain the $A^{t\bar{t}}_{FB}$ asymmetry observed at CDF. 

\begin{figure}[t]
\begin{center}
\vspace*{3mm}
\includegraphics[width=0.6\textwidth]{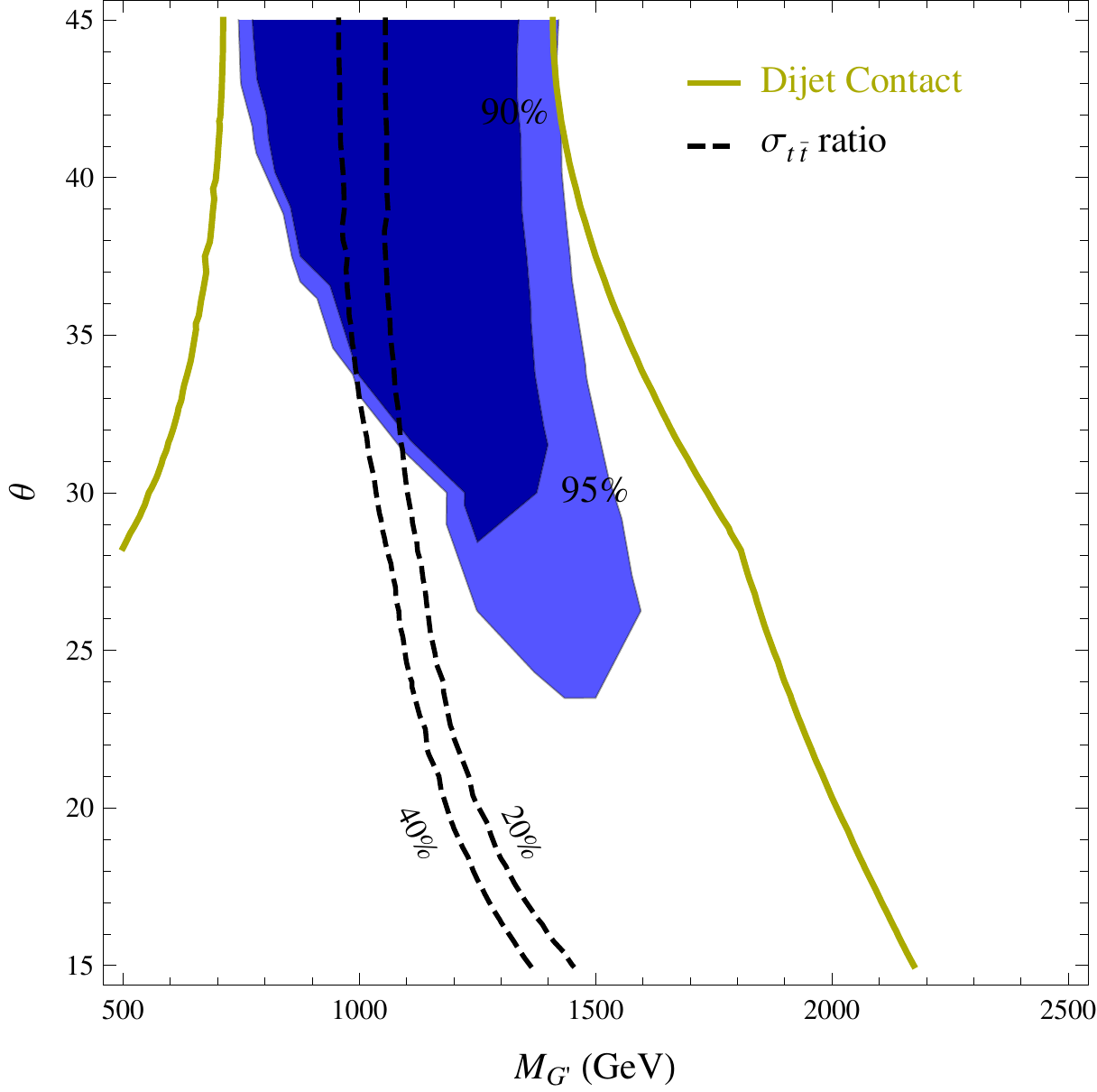} 
\caption{The same as Fig.~\ref{fig:twosite} but for $\Gamma_{G^\prime}/M_{G^\prime}\geq 20\%$. The dijet resonance searches no longer constrain this model.}
\label{fig:twositebroad}
\end{center}
\end{figure}

One may wonder whether enlarging the width of the axigluon can help to evade these contact interaction constraints. There are many ways to make the axigluon 
broader by forcing the $G^\prime$ to decay into other modes. For instance, in Ref.~\cite{Bai:2010dj} the axigluon $G^\prime$ can decay into two 
scalar color-octets (if they are lighter than $G^\prime$), which belong to the other components of the $\Sigma$ field. The scalar octet may further 
decay into two jets, so there are then four jets in the final state and the direct dijet searches become much less of a constraint. Without specifying 
a specific model, we study the effects of changing the $G^\prime$ width on the resonant dijet searches in Fig.~\ref{fig:twositebroad}. 
In Fig.~\ref{fig:twositebroad}, we show the corresponding fitted results assuming the width/mass ratio of $G^\prime$ to be greater than 20\% [if the width from 
Eq.~(\ref{eq:width}) is greater than 20\% of the mass, we use the width in Eq.~(\ref{eq:width})]. Although the narrow 
resonance searches do not apply in this case, the dijet contact interaction  searches still exclude all of the $A^{t\bar{t}}_{FB}$ preferred 
parameter space region. Therefore, we must consider other extensions of the two-site axigluon model.

\subsection{A Simplified Three-Site Model}
\label{sec:threesites}

One way to extend the minimal single axigluon model is to include an additional color-octet gauge boson. In Ref.~\cite{Chivukula:2010fk}, the authors compared 
the two-site model and the three-site model and concluded that the three-site model can not improve the fit to $A^{t\bar{t}}_{FB}$. Their 
arguments are based on the contact operator  analysis obtained by integrating out heavy gauge bosons, which can be justified only when the heavy gauge 
bosons have narrow widths and have masses above the collider limit.  However, one can see that the heavy gauge boson $G^\prime$ can be accessible at 
Tevatron from Fig.~\ref{fig:twosite} and that the width can be fairly large, so it is worthwhile to briefly revisit the three site model.

In this section we will reconsider the three-site model with 
the gauge group $SU(3)_1\times SU(3)_2 \times SU(3)_3$ plus two sigma fields $\Sigma_1$ and $\Sigma_2$, which transform as 
$(3, \bar{3})$ under $SU(3)_1\times SU(3)_2$ and $(\bar{3}, 3)$ under $SU(3)_2\times SU(3)_3$. For simplicity, we will assume that the gauge couplings 
of $SU(3)_1$ and $SU(3)_3$ are identical and are given by $h_1$, with the corresponding coupling of $SU(3)_2$ being $h_2$. We will further require 
that $\langle \Sigma_1 \rangle = \langle \Sigma_2 \rangle$, so there exists an {\em ad-hoc} $\mathbb{Z}_2$ symmetry in the gauge sector via the 
interchange of site ``1" and site ``3". Diagonalizing the gauge boson mass matrix, we determine the three mass eigenstates to be
\beqa
G^\mu &=& -\frac{\sin{\theta}}{\sqrt{2}}\,G^\mu_1 \,+\, \cos{\theta}\,G^\mu_2 \,-\, \frac{\sin{\theta}}{\sqrt{2}}\,G^\mu_3\,, \\
G^{\prime\,\mu} &=& \frac{1}{\sqrt{2}}\,G^\mu_1 \,-\, \frac{1}{\sqrt{2}}\,G^\mu_3 \,,\\
G^{\prime\prime\,\mu} &=& \frac{\cos{\theta}}{\sqrt{2}}\,G^\mu_1 \,+\, \sin{\theta}\,G^\mu_2 \,+\, \frac{\cos{\theta}}{\sqrt{2}}\,G^\mu_3 \,.
\eeqa
Here, the mixing angle is defined as $\tan{\theta} \equiv \sqrt{2}\,h_2/h_1$ and the physical region is identified as $0\leq  \theta < 90^\circ$. The 
mass ratio of the two massive gauge bosons is determined to be
\beq
\frac{M_{G^{\prime\prime}}}{M_{G^{\prime}}} \, =\, \frac{1}{\cos{\theta}} \,.
\eeq

Assigning the quantum numbers to SM quarks: $q_L$, $t_R$, $b_R$ to be triplets of $SU(3)_1$ and $(t, b)_L$, $q_R$  triplets of $SU(3)_3$, we derive the 
couplings of $G^\prime_\mu$ to quarks to be (the $\mathbb{Z}_2$ symmetry is explicitly broken by the quark charge assignments, so that the 
$G^\prime_\mu$ is unstable)
\beqa
g^t_V = g^q_V = 0 \,, \qquad 
g^t_A = - g^q_A =  \frac{1}{\sin{\theta}} \,,
\eeqa
and the corresponding couplings of  $G^{\prime\prime}_\mu$ to quarks are 
\beq
h^t_V = h^q_V = - \frac{1}{\tan{\theta}} \,, \qquad 
h^t_A = - h^q_A = 0 \,.
\eeq
Based on those couplings, we can see that $G^\prime$ behaves as an axigluon (with pure axial-vector couplings) and $G^{\prime\prime}$ as a coloron 
(with pure vector couplings). 

Including the exchanges of all three gauge bosons in the $s$-channel, we find the differential production cross section to be given by 
\beqa
\frac{d\sigma^{q\bar{q}\rightarrow t \bar{t}}}{d\cos{\theta^*}} &=& \frac{\pi\,\alpha_s^2\,\beta}{9\,\hat{s}}
\left\{ 1 + c^2 + 4m^2 + \frac{2\hat{s} (\hat{s} - M_{G^\prime}^2)}{ (\hat{s} - M_{G^\prime}^2)^2 + M^2_{G^\prime} \Gamma_{G^\prime}^2}  
\left(2 g^q_A g^t_A c \right) \right. \nonumber \\
&& \hspace{-2cm} \left.
+\, \frac{2\hat{s} (\hat{s} - M_{G^{\prime\prime}}^2)}{ (\hat{s} - M_{G^{\prime\prime}}^2)^2 + M^2_{G^{\prime\prime}} \Gamma_{G^{\prime\prime}}^2}  
\left[ h^q_V h^t_V (1 + c^2 + 4m^2) \right]
+ \frac{\hat{s}^2 }{ (\hat{s} - M_{G^\prime}^2)^2 + M^2_{G^\prime} \Gamma_{G^\prime}^2}  \left[ (g^q_A g^t_A)^2 (1 + c^2 - 4m^2) \right]
\right. \nonumber \\
&& \hspace{-2cm} \left.
+\, \frac{\hat{s}^2 }{ (\hat{s} - M_{G^{\prime\prime}}^2)^2 + M^2_{G^{\prime\prime}} \Gamma_{G^{\prime\prime}}^2}  \left[ (h^q_V h^t_V)^2 (1 + c^2 + 4m^2) 
\right] 
\right. \nonumber \\
&& \hspace{-2cm} \left.
+\,  \frac{\hat{s}^2 [(\hat{s} - M_{G^\prime}^2)( \hat{s} - M_{G^{\prime\prime}}^2 ) +  M_{G^\prime} M_{G^{\prime\prime}} \Gamma_{G^\prime} 
\Gamma_{G^{\prime\prime}} ] }{ [(\hat{s} - M_{G^\prime}^2)^2 + M^2_{G^\prime} \Gamma_{G^\prime}^2][ (\hat{s} - M_{G^{\prime\prime}}^2)^2 + 
M^2_{G^{\prime\prime}} \Gamma_{G^{\prime\prime}}^2 ]} 
  \left(4 g^q_A g^t_A  h^q_V h^t_V c \right)
\right\} \,.
\eeqa
Here, $\beta = \sqrt{1- 4m^2}$ is the velocity of the top quark, with $m = m_t /\sqrt{\hat{s}}$, and $c = \beta \cos{\theta^*}$ with $\theta^*$ as 
the production angle of $t$ in the center of mass frame of $t\bar{t}$ system.

\begin{figure}[t]
\begin{center}
\vspace*{3mm}
\includegraphics[width=0.6\textwidth]{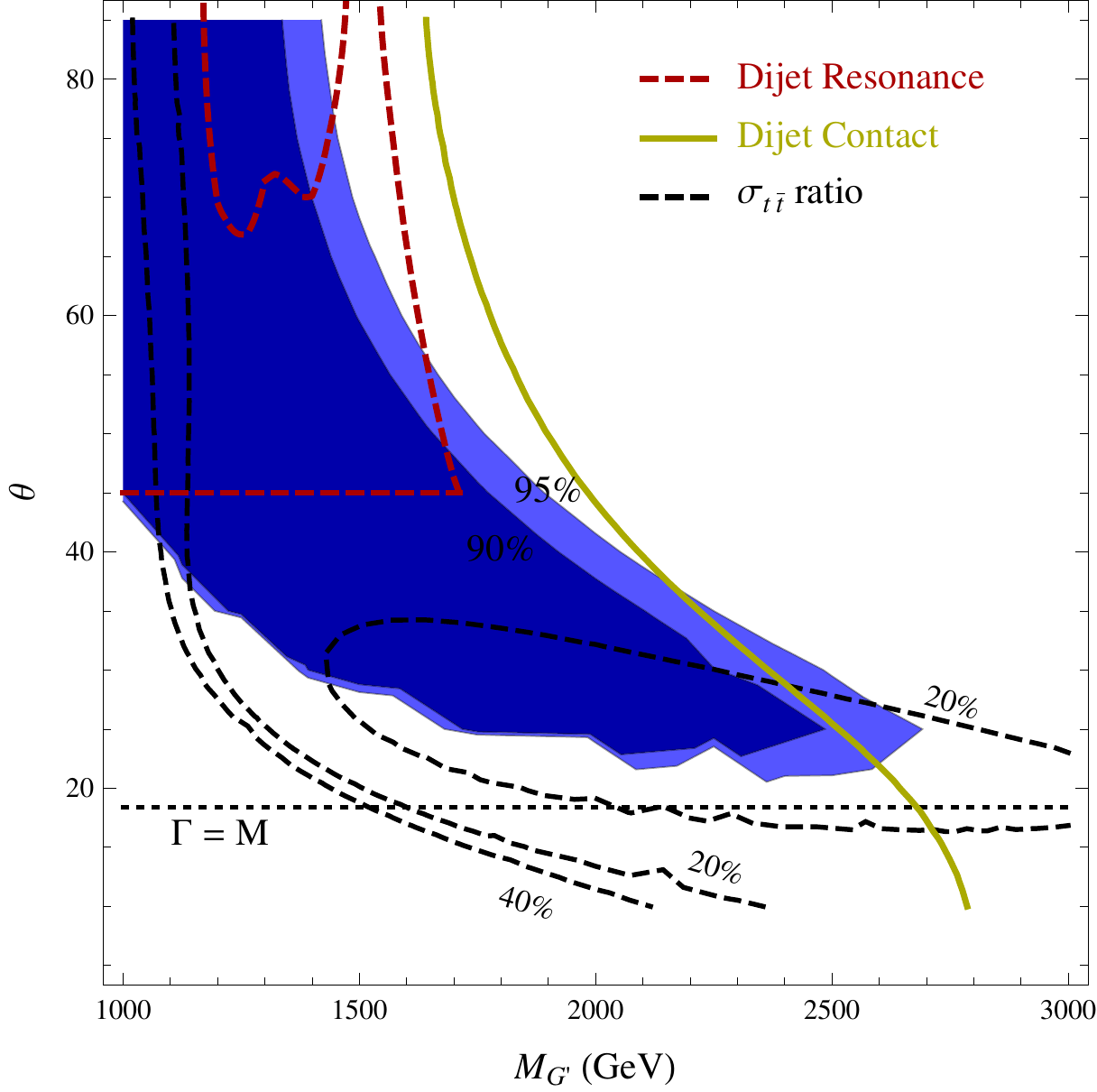} 
\caption{The fit of the simplified three-site model to the observed $A^{t\bar{t}}_{FB}$ at CDF. The best fit point has $\chi^2/d.o.f. = 4.4/2$ at 
$M_{G^\prime}=1.3$~TeV and $\theta=45^\circ$. Regions enclosed by the red dashed lines are excluded by the dijet narrow resonance searches. The region 
to the left side of the solid dark yellow line is ruled out by the search for dijet contact interactions. The $t\bar{t}$ production cross  section itself does not further constrain the relevant parameter space.}
\label{fig:threesite}
\end{center}
\end{figure}

Similar to the minimal two-site model, there are only two parameters in this model: $M_{G^\prime}$ and $\theta$. The corresponding 90\% and 95\% C.L. 
contour plot is shown in Fig.~\ref{fig:threesite}. 
One can see from this figure that although there exists a large region that is not ruled out by the dijet narrow resonance searches, however the search for  
contact interactions in dijets indeed excludes almost all of the parameter space. 

\subsection{A Two-Site Model with a New Vector-like Quark}
\label{sec:vectorfermion}

All of the previous models were highly constrained by the dijet searches, especially by the search of dijet contact  interactions.  So in this 
section we consider a model which can relax the dijet constraints without reducing the top quark forward-backward asymmetry.  We first notice 
that $A^{t\bar{t}}_{FB}$ only depends on the product of $g^q_A$ and $g^t_A$ for the two-site model while the dijet production cross section is only 
sensitive to the coupling $g^q_A$. If the relation between $g^q_A$ and $g^t_A$ as in Eq.~(\ref{eq:twositecoupling}) of the minimal two-site model 
can be broken such that $g^q_A$ can be treated independently from $g^t_A$, one then has the freedom to relax the dijet constraints. 

Returning to the gauge sector of the two-site model considered in Section~\ref{sec:twosites}, we now add a new vector-like (under SM gauge group) 
fermion, $\psi_{L,R}$, which are $SU(2)_W$ singlets and have charge $2/3$ under $U(1)_Y$. Under the extended gauge group, we assign 
$q_L$, $q_R$, $t_R$, $b_R$, $\psi_L$ to be triplets of $SU(3)_1$ and $(t, b)_L$, $\psi_R$ to be triplets of $SU(3)_2$. In the up-type quark sector, the 
general $4\times 4$ mass matrix can be diagonalized by a bi-unitary transformation acting on the left-handed and right-handed quarks. Since the 
left-handed mixing matrix is more highly constrained by the flavor observables, the right-handed mixing matrix has less serious constraints except for 
the mixing in the first two generations. Furthermore, because the first two-generation quarks have the same couplings under the new (flavor-changing) 
gauge boson $G^\prime$, the mixing between the right-handed quarks in the first two-generation is suppressed. To simplify our discussion, we assume the 
mixing matrix for the left-handed quarks to be approximately diagonal (we will discuss this assumption further in Section~\ref{sec:flavor}) and  introduce 
only one new mixing angle in the right-handed quark mixing matrix such that the new vector-like fermion only mixes with 
$u^{(f)}_R$ in the flavor basis.

The transition from the flavor basis to the mass eigenstate is then parametrized as
\beqa
u^{(m)}_R &=&  \cos{\alpha} \,  u^{(f)}_R  \,+\, \sin{\alpha}\, \psi^{(f)}_R\,, \\
\psi^{(m)}_R &=&  -\sin{\alpha} \,  u^{(f)}_R   \,+\, \cos{\alpha}\, \psi^{(f)}_R\,. 
\eeqa
The couplings of the axigluon $G^\prime_\mu$ to the various quarks in the mass eigenstate basis are found to be
\beqa
&& g_V^{(d, s, b, c)} = - \tan{\theta} \,, \quad g_A^{(d, s, b, c)} = 0 \,, \quad\quad g_V^t = \frac{1}{\tan{2\theta}}\,,\quad g_A^t = 
\frac{1}{\sin{2\theta}} \,, \nonumber \\
&& g_V^{u} =  - \tan{\theta} \,+\, \frac{\sin^2{\alpha}}{\sin{2\theta}} \,, \quad
     g_A^{u} = -  \frac{\sin^2{\alpha}}{\sin{2\theta}} \,.
 \label{eq:vectorfermioncoupling}
\eeqa
When $\alpha=\pi/2$, the couplings of the up and top quarks become identical to those in the minimal two-site model. From the above equation, one 
can see that the axial-vector couplings of the up and top quarks are different and one now has the freedom to increase $g_A^t$ by reducing $\theta$ 
and to simultaneously decease $g_{V, A}^u$ at the same time by choosing $\sin^2{\alpha}$ smaller than $\sin{2\theta}$. The total width of the 
$G^\prime_\mu$ in this model is then 
\beq
\Gamma(G^\prime) = \frac{\alpha_s\,M_{G^\prime}}{6}\,\frac{\left[(\cos^4{\alpha}+10) \tan^2{\theta} + (\sin^4\alpha + 1) \cot^2{\theta} - 2 
\sin^2\alpha \cos^2 \alpha \right] }{2}   \,.
\label{eq:width_vector}
\eeq

For simplicity, we only introduce one vector-like fermion which mixes solely with the up quark. In principle, other vector-like fermions may also be present which mix with the charm and top quarks. Such a new quark that mixes with top can be assigned the same quantum numbers as the top, so that the $G^\prime$ coupling to the top quark remains unchanged. A new quark that mixes with charm can have the same quantum numbers as the new quark that mixes with the up quark. To evade potential flavor constraints, one can impose an $SU(2)$ global symmetry, under which $(u_R, c_R)$ behave as an doublet and the two vector-like fermions as another doublet. For this choice, a mixing angle which is identical to $\alpha$ appears in the charm sector. The couplings $g_{A, V}^{c}$ will then be different from those in Eq.~(\ref{eq:vectorfermioncoupling}) and the width of $G^\prime$ will be modifed; however, the general arguments in this section will be untouched.

Before we perform a general fit for this model to the $A^{t\bar{t}}_{FB}$ data, we note that there exists parameter space regions where there is no new 
physics contribution to the dominant $t\bar{t}$ production process at the Tevatron:  $u\bar{u} \rightarrow t\bar{t}$, to the leading order in 
$\hat{s}/M^2_{G^\prime}$. This can happen when $\sin{\alpha} = \sqrt{2} \sin{\theta}$ for $0 \leq \theta \leq \pi/4$, which leads to $g_V^u =0$. Thus  
the contribution to the production cross section from the interference between the QCD gluon and the axigluon vanishes because it is proportional to 
the product $g_V^u g_V^t$. Let's first focus on this part of parameter space. The gauge couplings are then
\beqa
&& g_V^{(d, s, b, c)} =  g_A^{u} = - \tan{\theta} \,, \quad g_A^{(d, s, b, c)} = g_V^{u} = 0 \,, \quad\quad g_V^t = \frac{1}{\tan{2\theta}}\,,
\quad g_A^t = \frac{1}{\sin{2\theta}} \,.
\eeqa
The $G^\prime_\mu$  width is then simply 
\beq
\Gamma(G^\prime) = \frac{\alpha_s\,M_{G^\prime}}{6}\,\frac{\left( 11 \tan^2{\theta} + \cot^2{\theta}  \right) }{2}   \,.
\label{eq:width_vector2}
\eeq
\begin{figure}[t]
\begin{center}
\vspace*{3mm}
\includegraphics[width=0.55\textwidth]{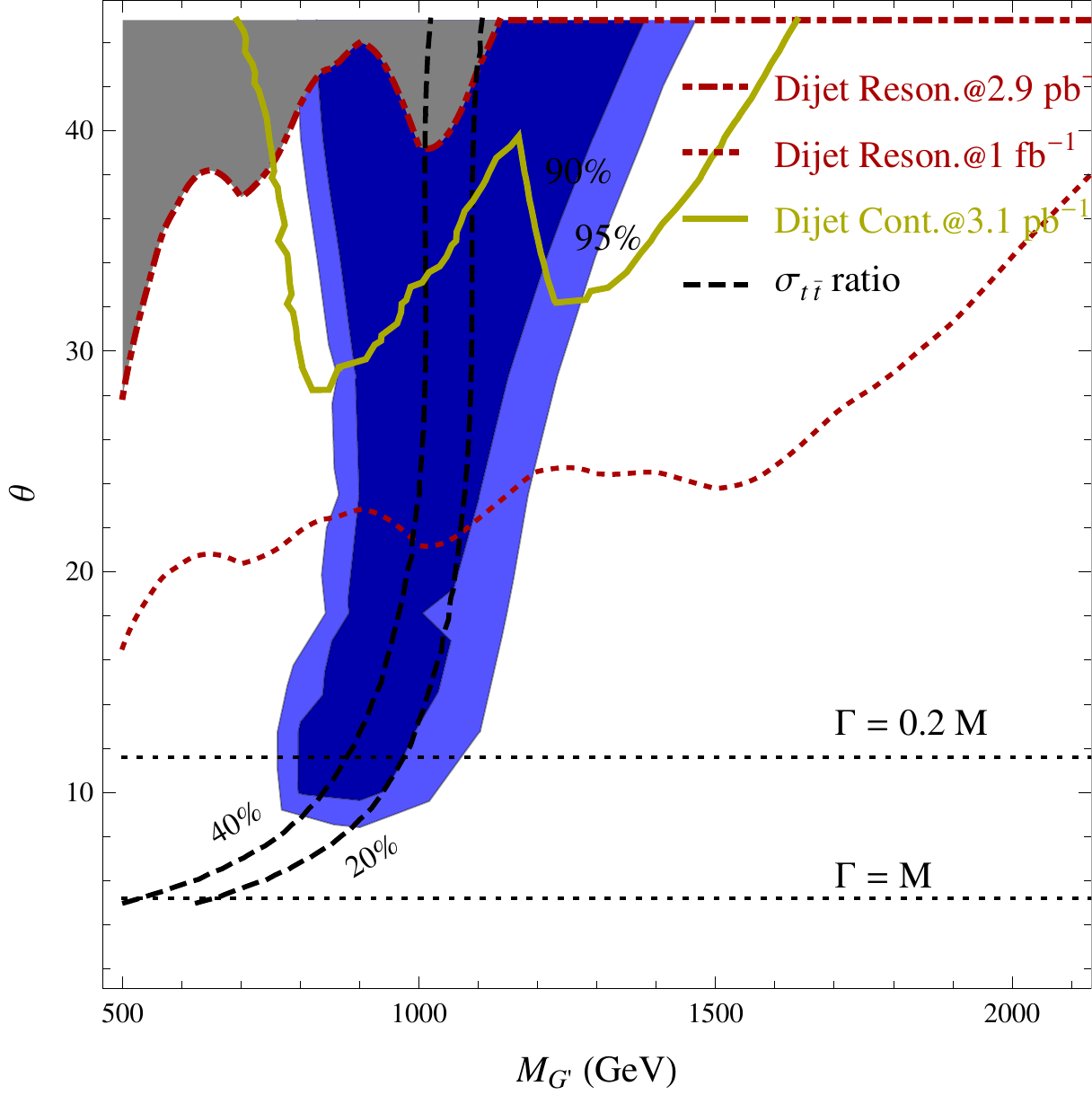} 
\caption{The fit for the model with one axigluon plus one additional vector-like fermion (Section \ref{sec:vectorfermion}) to the observed $A^{t\bar{t}}_{FB}$ at CDF. The best fit 
point has $\chi^2/d.o.f. = 5.3/2$ at $M_{G^\prime}=1.05$~TeV and $\theta=45^\circ$. The region above the red dot-dashed line is excluded at 95\% C.L. 
by the dijet narrow resonance search, while the region above the dark yellow solid line is also excluded at 95\% C.L. by the search of dijet contact 
operator interactions. The projected exclusion limit from the dijet narrow resonance search at the $7$ TeV LHC with 1 fb$^{-1}$ is shown by the red dotted line. 
The black dashed lines designate the regions (above and to the left) where the given percentage of the $t\bar{t}$ production cross section for $m_{t\bar{t}} > 450$~GeV arises 
from new physics.}
\label{fig:twositefermion1}
\end{center}
\end{figure}

The results of the fit are shown in Fig.~\ref{fig:twositefermion1}, where the red dotdashed line is the constraint from the current dijet narrow 
resonance searches at CMS with $2.9$~fb$^{-1}$, while the red dotted line is the projected exclusion limit for 1 fb$^{-1}$ at the LHC. The region 
above the dark yellow solid line is excluded by the search of dijet contact  interactions from ATLAS with 3.1 pb$^{-1}$. Although there is no 
modification of the $t\bar{t}$ production cross section from the QCD and $G^\prime$ interference term, the $t\bar{t}$ production cross section is 
modified by the new-physics-only contribution and has the relative contribution with respect to the SM leading-order production cross section shown 
by the black dashed lines. 
From this figure, we conclude that in the two-site plus one vector-like fermion model there 
exist a small region of parameter space which is allowed by all of the constraints. As an example, we show the fit results of $A^{t\bar t}_{FB}$ in Table~\ref{table:fitfermionmodel} for $M_{G^\prime} = 1.1$~TeV and  $\theta =30^\circ$.  
\begin{table}
\renewcommand{\arraystretch}{1.5}
\center
\begin{tabular}{| c | c | c |} 
\hline  
Selection & $M_{t \bar t} < 450$ GeV  &  $M_{t \bar t} > 450$ GeV \\ 
\hline  
Parton Level Exp. Data & $-0.116 \pm 0.146 \pm 0.047$ & $0.475 \pm 0.101 \pm 0.049$  \\ 
\hline 
Model Prediction & $0.06$ & $0.21$  \\ 
\hline 
\hline 
Selection & $| \Delta y | < 1.0$ & $| \Delta y | > 1.0$ \\ 
\hline  
Parton Level Exp. Data & $0.026 \pm 0.104 \pm 0.056$ & $0.611 \pm 0.210 \pm 0.147$ \\ 
\hline 
Model Prediction & $0.08$ & $0.27$  \\ 
\hline
\end{tabular}
\caption{The comparison of theoretic predictions and measured values for the two-site model with a new vector-like fermion. The $G^\prime$ mass is 1.1 TeV and the mixing angle is $30^\circ$. The total $\chi^2/d.o.f$ is 8.3/2.}
\label{table:fitfermionmodel}
\end{table}

From the contours of Fig.~\ref{fig:twositefermion1}, one can see that for a small mixing angle the best-fit region is insensitive to its exact value. 
This is easily understood since the product of the axial-vector couplings $g^u_A g^t_A = -1/(2\cos^2{\theta})$ is insensitive to $\theta$ for small 
$\theta$.  In this figure, we also show two horizontal, dotted black lines with different width/mass ratios to show that in the best-fit region  
$G^\prime$  is a narrow resonance. One can see that a large fraction of the parameter space will be covered with increasing luminosity at the LHC.

\begin{figure}[t]
\begin{center}
\vspace*{3mm}
\includegraphics[width=0.45\textwidth]{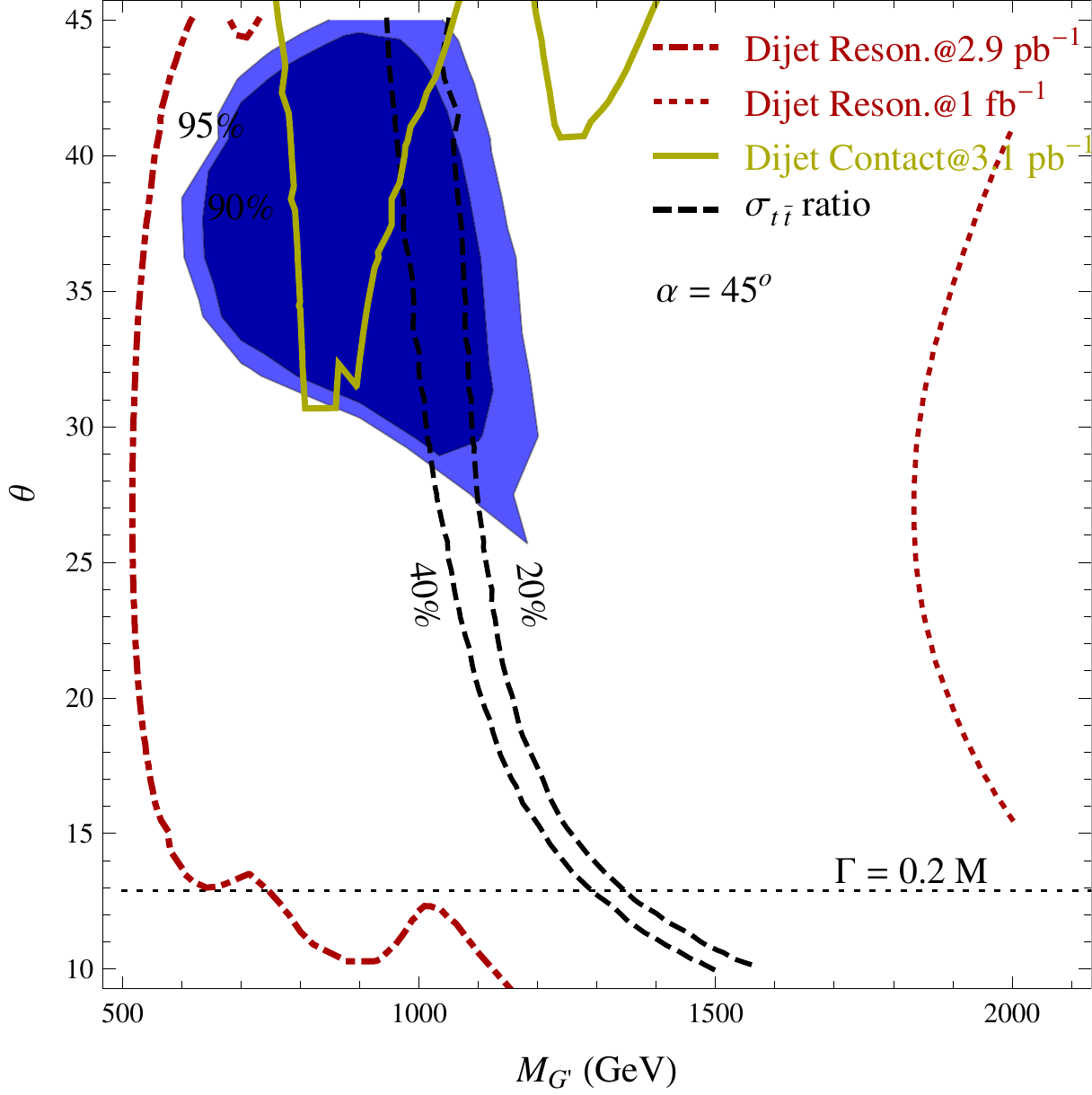} \hspace{1cm}
\includegraphics[width=0.45\textwidth]{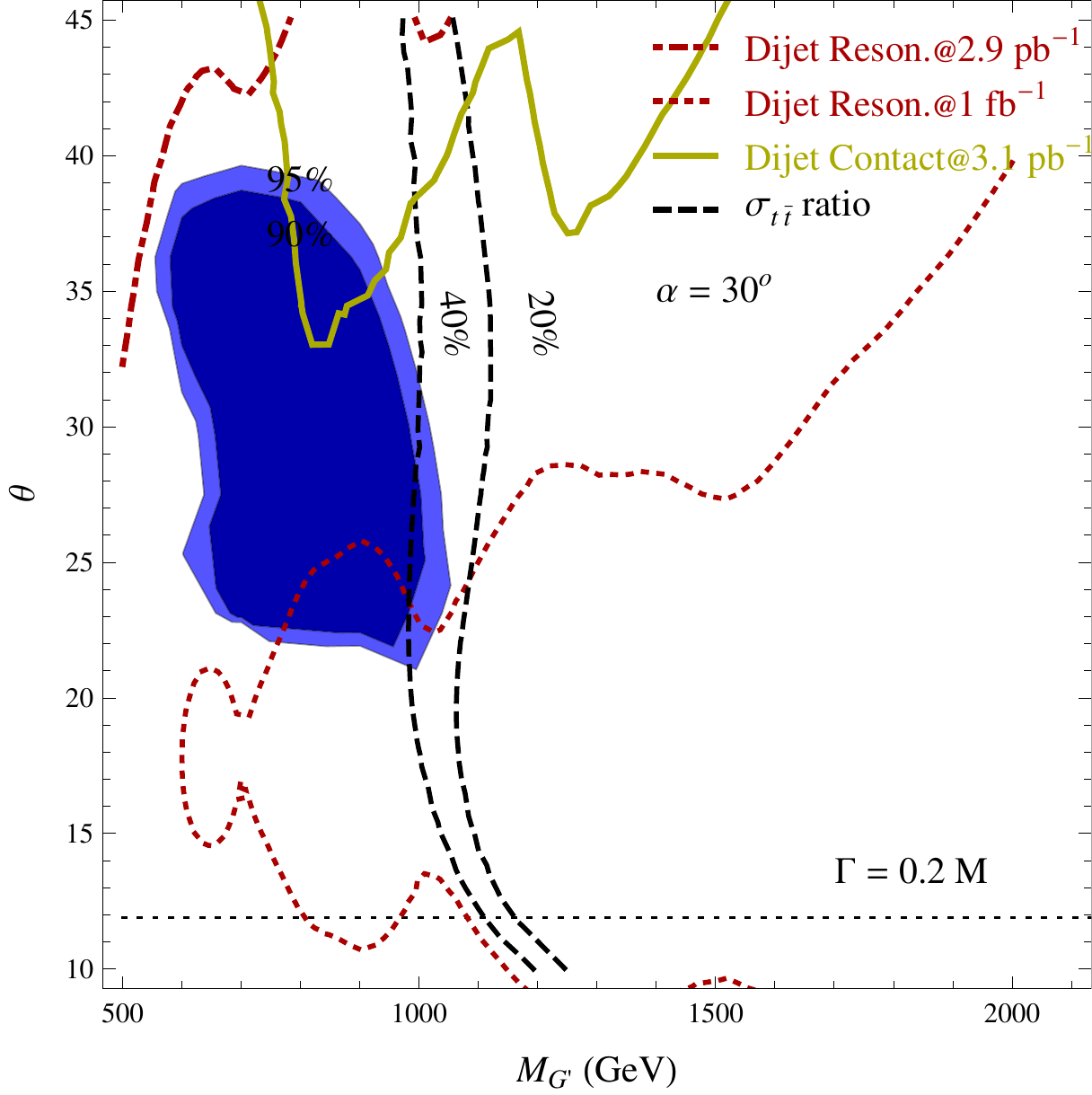} 
\caption{Left panel: the fit of the model with one axigluon plus one additional vector-like fermion (sec.~\ref{sec:vectorfermion}) to the observed $A^{t\bar{t}}_{FB}$ at CDF with 
$\alpha = 45^\circ$. The best fit point has $\chi^2/d.o.f. = 4.4/2$ at $M_{G^\prime}=849$~GeV and $\theta=42^\circ$. The region to the left of the 
red dotdashed 
line is excluded by the current dijet narrow resonance searches with luminosity 2.9 pb$^{-1}$. The region above the dark yellow solid line is excluded 
by the search for dijet contact interactions. Right panel: same as the left one but now for $\alpha = 30^\circ$. The best fit point has a  
$\chi^2/d.o.f. = 2.6/2$ lying at $M_{G^\prime}=824$~GeV and $\theta=29^\circ$.}
\label{fig:twositefermion2}
\end{center}
\end{figure}

For more general values of  $\alpha$, we show the contour regions in Fig.~\ref{fig:twositefermion2} for $\alpha = 45^\circ$ and $\alpha = 30^\circ$, respectively. 
In those two figures, the QCD and $G^\prime$ interference term also contributes to the $t\bar{t}$ production cross section, so the parameter space is 
highly constrained by $t\bar{t}$ resonance searches.
Comparing the two plots in Fig.~\ref{fig:twositefermion2}, we can see that for a smaller $\alpha$ a lighter $G^\prime$ is preferred by the fit to 
$A^{t\bar{t}}_{FB}$ because the product $g_A^u g_A^t$ is reduced. As a result, the parameter space for a smaller $\alpha$ leads to a larger $t\bar{t}$ 
production cross section and is thus less favored. For $\alpha = 30^\circ$, the best fit is $\chi/d.o.f. = 2.6/2$ at $M_{G^\prime}=824$~GeV and 
$\theta=29^\circ$. For this point, we obtain $A^{t\bar{t}}_{FB}=(0.14, 0.53)$ for the two different rapidity bins and $(0.06, 0.42)$ for the two bins in 
invariant masses.

\section{Flavor Constraints}
\label{sec:flavor}
To reproduce the measured $t \bar t$ forward-backward asymmetry, the coupling of $G^\prime$ to the top quark is forced to have the opposite sign as the 
coupling of $G^\prime$ to the $u$ and $d$ quarks.  This necessitates flavor violation at some level, because it means that the $G^\prime$ coupling cannot be 
proportional to the identity matrix in flavor space.  Here, we compute the size of these flavor effects.

In order to determine the tree-level contributions of a heavy spin-1 octet, $G^\prime$, to FCNC processes, we must first recall how they arise 
in a general theory {\cite{Hewett:1988xc}}. Let $D^{0\,T}=(d,s,b)^0$ and $U^{0\,T}=(u,c,t)^0$ represent the three SM generations of $Q=-1/3$ and 
$Q=2/3$ quarks in the weak eigenstate basis, respectively. Spontaneous symmetry breaking via the Higgs Yukawa couplings then generates a mass 
matrix for these fields in the form $\bar D_L^0 M^d D_R^0 +h.c.$ (and similarly for the up-type quarks). In order to diagonalize this 
matrix, $M^d$, one needs to introduce the bi-unitary transformations $D_{L,R}=U^D_{L,R}D_{L,R}^0$ so that $U^D_L M^d U_R^{D\,\dagger} =M_D^d$ where $M_D^d$ 
is the $Q=-1/3$ diagonal mass matrix in the mass eigenstate basis. Note that {\emph{if}} $M^d$ is Hermitian, we are then free to choose 
$U^D_L=U^D_R=U^D$. Note further that in this notation the conventional CKM matrix is given by a product of LH-rotations for the up-type and down-type quarks: 
$V_{CKM}=U_L^U U_{L}^{D\,\dagger}$ and that, in the SM, the RH-rotations play no role.  

In what follows we will first compute the contribution of $G'$ to $B_q$-$\bar B_q$ mixing (with $q=d,s$), noting that the analysis for $D$-$\bar D$ mixing is identical with the substitutions $b\to c$ and $q\to u$.  It is conceptually important to consider both types of mesons, because constraints from either alone might be evaded by pushing all the quark mixings into either the up or down-type sectors.  We will also compute the constraints from $K$-$\bar K$ mixing via a direct matching to the operator coefficient limits given by the UTfit collaboration \cite{Bona:2007vi}.  We use a different analysis for $K$-$\bar K$ because the strange quark is not heavy.

The interactions of the field $G^\prime$ with the $Q=-1/3$ fermions in the weak interaction basis are
\begin{equation}
{\cal L}=-g_s \bar D_L^0 A_Lt^a \gamma^\mu D_L^0 \,G^{\prime\, a}_\mu +(L\to R)\,, 
\end{equation}
where $t^a$ are the color generators and the matrices $A_{L(R)}=diag(a_1,a_2,a_3)_{L(R)}$ are the explicit couplings in this basis.  Note that the $A$ matrices are 
bi-fundamentals under $SU(3)$ flavor symmetries, and as such, they are a new source of flavor violation. Rotating to the mass eigenstate basis, we have new couplings that take the explicit form
\begin{equation}
C^D_{L, kj}= (U_L^D)_{km} a_{Lm} \delta_{mi} (U_{L}^{D\,\dagger})_{ij}\,, 
\end{equation}
and similarly for $L\to R$. Here, we are specifically interested in the case of {\it off-diagonal} couplings, $k\neq j$ with $(a_1=a_2\neq a_3)_{L(R)}$ 
as we are assuming that the first two generations have identical couplings.  Since by unitarity  $(U_{L}^D)_{ki}(U_{L}^{D\,\dagger})_{ij}=\delta_{kj}$, and similarly for the right-handed and up-type sectors,
we find the off-diagonal couplings to be given by
\begin{equation}
C^D_{L, kj}= (a_3-a_1)_L (U_L^D)_{k3}(U_L^{D\,\dagger})_{3j}=(a_3-a_1)_L (U_L^D)_{k3}(U_L^{D\,*})_{j3}~~k\neq j\,, 
\end{equation}
and similarly for $L\to R$. Of course, all of this naturally repeats itself in the $Q=2/3$ sector, with $U_{L}^D \to U_{L}^U$.

Now consider the FCNC contributions from $G^\prime$ to $B_q$-$\bar B_q$ mixing; the effective Lagrangian is
\begin{equation}
{\cal L_{\mathrm {FCNC}}}=-g_s \bar b\, t^a \gamma^\mu (C_{L, 3q}^D \,P_L+C_{R, 3q}^D\, P_R)q\,G_a^{\prime\,\mu}+h.c.\,.
\end{equation}
To obtain our results we follow the discussion as given in Section 4.2 of Ref.~{\cite{Blanke:2008zb}} which contains a very similar set-up. The procedure 
is, essentially, as follows: ($i$) draw the contributing diagrams arising from ${\cal L_{\rm FCNC}}$ and integrate $G^\prime$ out. ($ii$) Contract the color 
structure and Fierz transform the product of the two currents to express the resulting expression in terms of a known basis of 4-fermion 
operators \cite{Golowich:2007ka}. ($iii$) Compute the renormalization group running of these operators down to the scale of the meson masses, and compute the appropriate matrix elements. 
Following Ref.~\cite{Blanke:2008zb}, the effective Hamiltonian is given by
\begin{equation}
{\cal H}_{\Delta B=2}={{2\pi \alpha_s}\over {3M_{G^\prime}^2}} \Big [(C_{L,3q}^{D})^2 Q_{1L}+ (C_{R, 3q}^{D})^2 Q_{1R}-C_{L,3q}^D C_{R}^D Q_2-6\,C_{L,3q}^D C_{R,3q}^D Q_3\Big]\,, 
\end{equation}
with the operators
\bea
Q_{1 L,R} & = & (\bar q_{L,R} \gamma_\mu b_{L,R}) (\bar q_{L,R} \gamma^\mu b_{L,R}) \,, \\
Q_{2} & = & (\bar q_{L} \gamma_\mu b_{L})(\bar q_{R} \gamma_\mu b_{R})  \,,\\
Q_{3} & = & (\bar q_L b_R) (\bar q_R b_L) \,.
\eea
This Hamiltonian describes $B$, $D$, and $K$ meson mixing with appropriate replacements.  

To proceed in the case of $B$-$\bar B$ mixing, we compute the matrix elements of ${\cal H}_{\Delta B=2}$ and compare to bounds from the UTfit Collaboration  \cite{Bona:2007vi}.  We present the results in terms of the matrix element of
the LL operator $Q_{1L}$, which we write as $r_1 \langle Q_{1L} \rangle $, where $r_1\simeq 0.77$ and accounts for the renormalization group running of $Q_{1L}$ from the TeV scale to the $b$ quark mass, as is appropriate for $B$-$\bar B$ mixing ($r_1 \simeq 0.70$ for $D$-$\bar D$ mixing and accounts for the additional RG evolution to the charm quark mass).  The matrix elements of the other operators differ by factors that we write as $R_{2,3}$, which account for their RG running, so in total we find
\begin{equation}
\langle B | {\cal H}_{\Delta B=2} | \bar B \rangle = {{2\pi \alpha_s}\over {3M_{G'}^2}} ~\Big [(C_{L,3q}^D)^2+(C_{R,3q}^D)^2 -C_{L,3q}^D C_{R,3q}^D R_2 - 6\,C_{L,3q}^D C_{R,3q}^D R_3\Big]\,,
\end{equation}
with $R_{2,3}$ given by the expressions in \cite{Golowich:2007ka} such that (taking $N_c=3$)
\begin{equation}
R_2={{1}\over {\sqrt r_1}} \Big( -{{3}\over {4}} -{{1}\over {2}} ~{{M_B^2}\over {(m_b+m_q)^2}}\Big),~~~R_3={{1}\over {r_1^5}} \Big({{1}\over {8}}+
{{3}\over {4}} ~{{M_B^2}\over {(m_b+m_q)^2}} \Big)\,.
\end{equation}
Numerically we obtain $R_2\simeq -1.7$ and $R_3\simeq 4.8$. The bound on $M_{G^\prime}$ is then obtained by comparing to the cutoff limit from UTfit \cite{Bona:2007vi}:
\begin{equation}
{{2\pi \alpha_s}\over {3M_{G'}^2}} ~\left|(C_{L,3q}^{D})^2+(C_{R,3q}^D)^2 -27 \,C_{L,3q}^D C_{R,3q}^D \right| < \frac{1} {\Lambda_B^2}\,,
\end{equation}
with $\Lambda_B = 210\,(30)$ TeV for $B_{d}$ ($B_s$) mixing.  This also allows a direct comparison to the analysis of Ref.~\cite{Chivukula:2010fk}, with which we disagree.   Numerically this reduces, in the case of $B_d$ mixing, to 
\begin{equation}
M_{G'} \gtrsim (100~{\rm {TeV}}) ~\left|(C_{L, 31}^{D})^2+(C_{R,31}^{D})^2 -27\,C_{L,31}^D C_{R,31}^D \right|^{1/2}\,.
\end{equation}
An identical analysis applied to the up-type sector for $D$-$\bar D$ mixing gives
\begin{equation}
{{2\pi \alpha_s}\over {3M_{G^\prime}^2}} ~\left|(C_{L,21}^U)^2+ (C_{R,21}^U)^2 - 60\, C_{L,21}^U C_{R,21}^U \right|^2 < {{1}\over {\Lambda_D^2}}\,,
\end{equation}
and numerically with $\Lambda_D = 1.2 \times 10^3$ TeV~\cite{Bona:2007vi} we find
\begin{equation}
M_{G'}  \gtrsim (600~{\rm {TeV}}) ~\left | (C_{L,21}^U)^2+ (C_{R,21}^U)^2 - 60\, C_{L,21}^U C_{R,21}^U \right|^{1/2}\,.
\end{equation}

In the case of $K$-$\bar K$ mixing we will directly compare the coefficients of these operators to the UTfit bounds  \cite{Bona:2007vi}.   From $Q_{1 L,R}$ we find that
\be
M_{G'}\gtrsim (500~{\rm {TeV}}) ~\left|(C_{L,21}^D)^2+ (C_{R,21}^D)^2\right|^{1/2}\,,
\ee
and from $Q_{3}$
\be
M_{G'} \gtrsim (5000~{\rm {TeV}}) ~|C_{L,21}^DC_{R,21}^D|^{1/2}\,.
\ee
These limits derive from the real part of the $K$-$\bar K$ mixing; the imaginary part gives limits that are stronger by a factor of $15$.

We cannot proceed without making model-dependent assumptions, so in particular, we cannot set a firm exclusion limit.  To give a reasonable picture of the situation, we will consider two very simple model  assumptions, where ($\alpha$) the quark mass matrices are Hermitian so that $U_{L}^D=U_{R}^D=U^D$ and $U_{L}^U=U_{R}^U=U^U$ and ($\beta$) where the right handed $U_{R}^U = U_{R}^D = \mathbb{I}_3$.  Furthermore, for convenience we assume that all three generations have identical vector couplings to $G^\prime$, while the first two generations and the third have distinct axial couplings. 

For the choice ($\alpha$) we obtain (with the couplings defined in units of $g_s$) 
\begin{equation}
C_{L,31}^D=-C_{R,31}^D= (g_A^t - g_A^q)\,U_{33}^D\, U_{13}^{D\,*} \ \ \ \mathrm{and} \ \ \ C_{L,21}^U=-C_{R,21}^U= (g_A^t - g_A^q)\, U_{23}^U \,U_{13}^{U\,*} \, ,
\end{equation}
note that because $G'$ couples universally to the first two generations, to induce an FCNC we must couple through the third generation.  The numerical bounds are listed in the second column of Table~\ref{table:contraintalpha}.
\begin{table}
\renewcommand{\arraystretch}{1.8}
\center
\begin{tabular}{| c | c || c |} 
\hline  
       & \multicolumn{2}{c|}{Lower bound on $M_{G^\prime}$ in TeV} \\ 
\hline
\hline  
   &   $(\alpha)$    &     $(\beta)$     \\
   \hline
Kaon \, System  &  $5000 \times |U_{23}^D \,U_{13}^{D\,*}|$  & $500 \times |U_{23}^D \,U_{13}^{D\,*}|$ \\ 
\hline  
$B$ \,  Mixing   &  $500 \times   |U_{33}^D\,U_{13}^{D\,*}|$  & $100 \times   |U_{33}^D\,U_{13}^{D\,*}|$ \\ 
\hline  
$D$ \,  Mixing   & $ 4500 \times |U_{23}^U\,U_{13}^{U\,*}|$  & $ 600 \times |U_{23}^U\,U_{13}^{U\,*}|$ \\
\hline
\end{tabular}
\caption{The constraints on the axigluon mass and mixing matrices from meson mixing for the two model choices discussed in the text. The common multiplicative factor $|g_A^t - g_A^q|$ is not included in the table.  The Kaon limit would be a factor of $15$ stronger in the presence of $O(1)$ phases.}
\label{table:contraintalpha}
\end{table}
Alternatively, for the choice $(\beta)$ we have
\begin{equation}
C_{L, kj}^{U, D}= (g_A^t - g_A^q)\, U^{U, D}_{k3} U^{U, D\,*}_{j3} \ \ \ \mathrm{and} \ \ \ C_{R, kj}^D = -C_{R, kj}^{U} = 0 \, ,
\end{equation}
and so we avoid the large RG enhancement, giving the weaker constraints listed in the third column of  Table~\ref{table:contraintalpha}.

Although our models determine the coupling $g_A^{t, q}$, without an accompanying model of flavor we cannot determine what values the matrix elements of $U$ take.  
All that we know are the elements of the CKM matrix, $V_{CKM}=U_{L}^U U^{D\,\dagger}_{L}$. In the general case all elements of the matrices $U_{L,R}^{U,D}$ 
could enter the observables, so we can no longer operate in a basis where, \eg, one of the original mass matrices is diagonal.  If we assume  
(without any particular justification) that $|U^D_{33}U^{D\,*}_{13}|=|V_{td}^*V_{tb}| \simeq 8.4\cdot 10^{-3}$ and $|U^D_{23}U^{D\,*}_{13}|=|V_{ts}^*V_{td}| \simeq 3.5\cdot 10^{-4}$  {~\cite{Nakamura:2010zzi}} 
then we find that  $M_{G'} \gtrsim 5 |g_A^t - g_A^q|$ TeV for the choice ($\alpha$), and  $M_{G'} \gtrsim  0.8 |g_A^t - g_A^q|$ TeV for choice ($\beta$).  Here we have assumed that all complex phases are small; if we assume ${\cal O}(1)$ phases then these constraints become about a factor of three stronger.  Note that in the extreme limit that $U_{L}^D =U_{R}^D = \mathbb{I}_3$ and $U_L^D =\mathbb{I}_3$ while $V_{CKM} = U_{L}^U$, $M_{G^\prime}$ and $g_A$ are completely unconstrained by considerations of flavor in the down sector.  Then the only constraint comes from 
$D$-$\bar D$ mixing, and we have $|U_{23}^U\,U_{13}^{U\,*}|=|V_{ts}^*V_{td}| \simeq 3.5 \cdot 10^{-4}$, giving $M_{G'} \gtrsim 1.6 |g_A^t - g_A^q|$ TeV for ($\alpha$), and $M_{G'} \gtrsim 200 |g_A^t - g_A^q|$ GeV for choice ($\beta$).  The latter constraint is trivially satisfied by all of our models.  

To summarize:  there are no model independent flavor constraints; generic axigluon models have significant tension with flavor, but there are plausible flavor scenarios where this tension disappears.

\section{Signatures at the LHC and Tevatron}
\label{sec:LHC}

The axigluon  explanation of the observed $t \bar t$ forward-backward asymmetry can be tested at the LHC in either the dijet or the $t\bar t$ 
final state. 
Almost all of the $A^{t\bar t}_{FB}$ preferred parameter space in the above models will be explored with 1 fb$^{-1}$ of data at the 7 or 8 TeV LHC. 

\subsection{Dijet Resonances}
\label{sec:dijet}

By rescaling the results of the CMS dijet narrow resonance search~\cite{Khachatryan:2010jd} performed with 2.9 pb$^{-1}$, we can determine how much luminosity will be needed 
to discover a given model.  For instance, for the model with one axigluon and one vectorlike fermion presented above with 
$M_{G^\prime} = 1.1$ TeV and a mixing angle of $\theta = 30^{\circ}$, the dijet production cross section is suppressed by $1/5$ relative 
to a standard ``coloron" model \cite{Hill:1991at}, so naively we expect that about 35 pb$^{-1}$ will be sufficient to exclude this model point at $95 \%$ C.L. On the other 
hand,  with a luminosity $\sim$ 250 pb$^{-1}$ the axigluon with this set of parameters will be discovered at the 7 TeV LHC. The projected exclusion limits with 1 fb$^{-1}$ of luminosity is presented in Fig. \ref{fig:twositefermion1}.  We see that if it exists, a new axigluon resonance 
that can explain the $t \bar t$ forward-backward asymmetry should be discovered sometime this year!

For an axigluon described by the phenomenological model presented in Fig.~\ref{fig:generalmodel}, its width should be greater than  
20\% of its mass and it will appear as a wide resonance. The dijet narrow resonance searches will likely no longer be applicable. One could then employ an analysis similar to the quark contact interaction  searches~\cite{Collaboration:2010eza} using the dijet rapidity difference to distinguish the signal events from the 
background. For  $M_{G^\prime} = 2 $~TeV, $g_A^u=1.5$, $g_A^{t} = -2$ and $g_V =0$, one can exclude this axigluon at 95\% C.L. with $\sim$ 44 pb$^{-1}$ 
of data assuming a 5\% systematic error on the SM backgrounds following the analysis presented in~\cite{Collaboration:2010eza}. However, in order to obtain a 
$5\sigma$ discovery, one needs to modify the strategy appearing in~\cite{Collaboration:2010eza} by choosing appropriate invariant mass bins and comparing the 
signal and background shapes in order to reduce this systematic error. We encourage the ATLAS and CMS collaborations to adapt their dijet searches in order to cover all 
the parameter space of the above phenomenological model.

\subsection{Top-Anti-Top Resonances}
\label{sec:ttbar}
As discussed above, the $t\bar t$ mass distribution is quite sensitive to the
presence of a new color-octet vector boson.  The large $t\bar t$ cross section \cite{Khachatryan:2010ez}
coupled with the increased mass reach make the LHC an ideal environment for such
resonance searches.  Here, we explore the reach of the
LHC in this observable for some $A^{t\bar t}_{FB}$-preferred and viable parameter points of the above axigluon models.

\begin{figure}[t]
\begin{center}
\vspace*{3mm}
\includegraphics[width=0.7\textwidth]{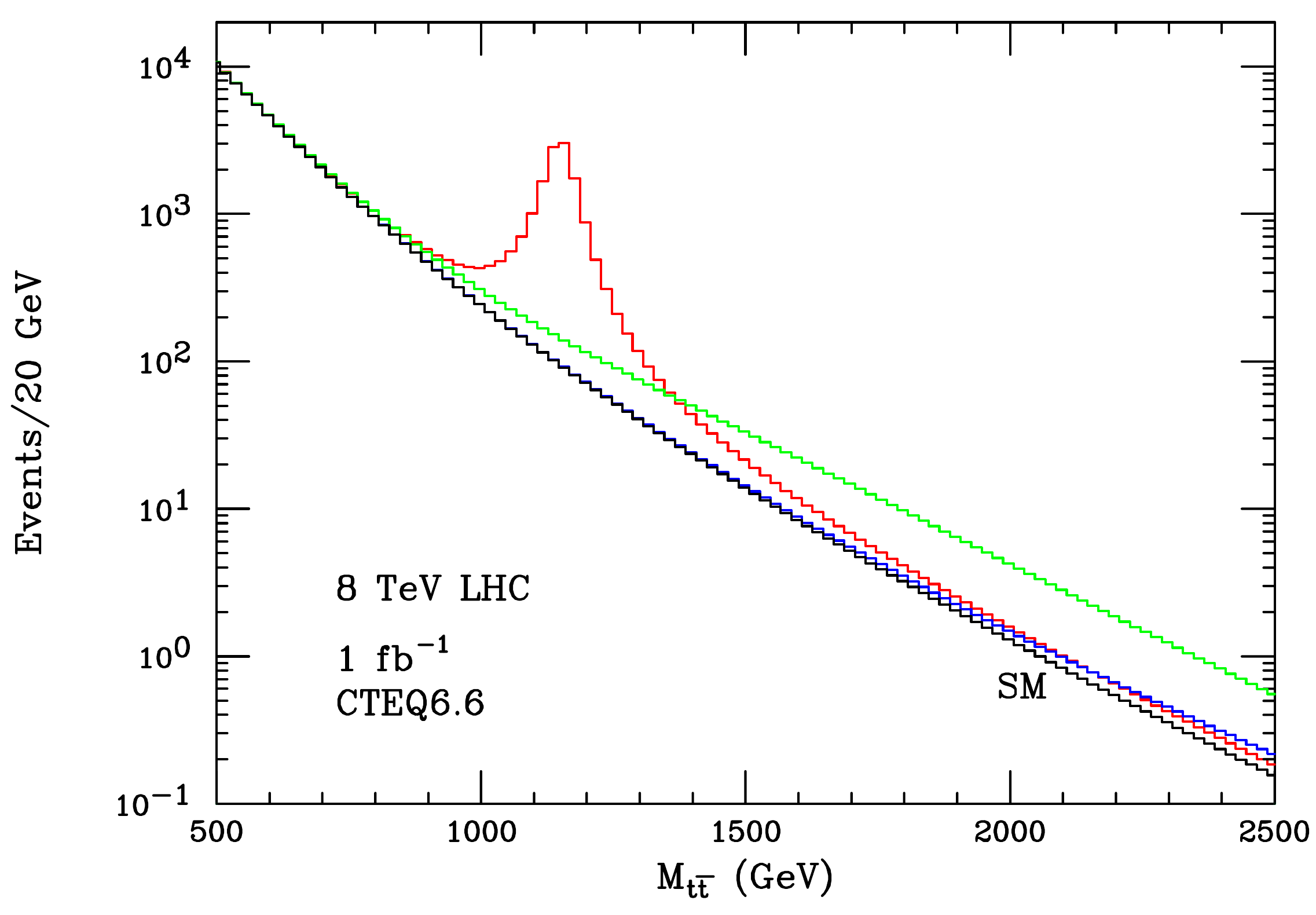} 
\caption{The $M_{t\bar t}$ distribution from the axigluon contribution plus the SM background at leading order. The red line is for the model with one 
axigluon and one vectorlike fermion described above with $M_{G^\prime}=1100$ GeV and $\theta=30^\circ$. The green(blue) line 
is for the phenomenological axigluon model with  $M_{G^\prime} = 1 (2)$~TeV, $g_A^u=1.5$, $g_A^{t} = -2$ and $g_V =0$.  The black line represents the SM.}
\label{fig:lhcmtt}
\end{center}
\end{figure}

The new axigluon, which does not couple to two gluons, is exchanged only in $s$-channel $q\bar q$ annihilation and so does not interfere with the gluon 
fusion contribution.  The signature is striking as shown in Fig.~\ref{fig:lhcmtt}.  
Here, the red line displays the $t\bar t$
invariant mass distribution at the 8 TeV LHC for the model with one axigluon and one vectorlike fermion discussed above, 
taking $M_{G^\prime}=1100$ GeV and $\theta=30^\circ$, while the black curve represents the SM. With 1 fb$^{-1}$ luminosity, it is clear that one can easily find this new resonance on top 
of the SM background.  The green (blue) line shows the contributions arising from the SM plus the phenomenological axigluon model with  
$M_{G^\prime} = 1 (2)$~TeV, $g_A^u=1.5$, $g_A^{t} = -2$ and $g_V =0$. Since $\Gamma/M \approx 28\%$, no obvious bump appears in the distribution and 
one needs to employ a contact interaction search in $t\bar t$ final states.

We next estimate the sensitivity of the 8 TeV LHC to the presence of these new states.  We model our analysis after the strategy detailed 
by the ATLAS Collaboration for generic $t\bar t$ resonance searches~\cite{Aad:2009wy}\cite{Cogneras2006}, including a new color octet boson from technicolor 
models.  We only consider the semi-leptonic
top-quark decays into muons and electrons (see Ref.~\cite{Barger:2006hm}\cite{Choudhury:2007ux}\cite{Baur:2008uv} for detailed studies with the 14 TeV LHC). Other channels, 
such as the dilepton channel~\cite{Bai:2008sk} can also be used to improve the discovery sensitivity.  The ATLAS studies find that the main
source of background for $t\bar t$ resonances originates from SM top pair production,
while reducible backgrounds such as $W+$jets are negligible for resonance
searches.  The resulting dominant background arises from combinatorics.
We employ the $M_{t\bar t}$ reconstruction efficiency presented in~\cite{Aad:2009wy}\cite{Cogneras2006},
which is at the level of a few percent and drops significantly for increasing $M_{t\bar t}$. We note that new techniques have recently been developed that reconstruct top-quarks produced with 
large enough transverse momentum such that their decay
products can be tagged as a single jet~\cite{Kaplan:2008ie}\cite{Plehn:2010st}\cite{Rehermann:2010vq}\cite{Abdesselam:2010pt}; however employing
these methods is beyond the scope of our simple analysis here.  We estimate the
combined uncertainties in the NLO $t\bar t$ cross section and the parton
distribution functions at high invariant masses to be of order 50\%~\cite{Frederix:2007gi}\cite{Moch:2008qy}. 

Requiring $S/\sqrt{B} \ge 5$ as a discovery criteria, and including both statistical and 50\% systematic theoretical errors, we find that for the model with one axigluon and one vectorlike fermion and taking  
$\theta = 30^\circ$, the axigluon can be discovered up to 1.7 TeV at the 8 TeV LHC with 1 fb$^{-1}$.   As the parameter $\theta$ is varied, we find that the axigluon cannot be observed at the level of $S/\sqrt B\geq 5$ for $\theta<13^\circ$.  For $\theta\geq 13^\circ$, the axigluon search reach is in the range $1100-1700$ GeV as $\theta$ increases in value.
Thus, the $A^{t\bar t}_{FB}$-preferred region in 
Fig.~\ref{fig:twositefermion1} with a narrow axigluon or at large $\theta$ will be probed by the early LHC data. For the phenomenological axigluon model 
with a large axigluon width presented in Fig.~\ref{fig:generalmodel}, a narrow resonance search is no longer applicable and a contact interaction analysis with $t\bar t$ final states must be performed.  However, it is clear from Fig. \ref{fig:lhcmtt} large luminosities or an increase in center-of-mass energy will be required to observe a $\sim 2$ TeV axigluon in the $t\bar t$ channel for this scenario. For such heavy states, their contribution to $M_{t\bar t}$ is of order the NLO and PDF uncertainties.

\subsection{Production in Association with a Weak Gauge Boson}
\label{sec:GprimePlusW}

\begin{figure}[t]
\begin{center}
\vspace*{3mm}
\includegraphics[width=0.6\textwidth]{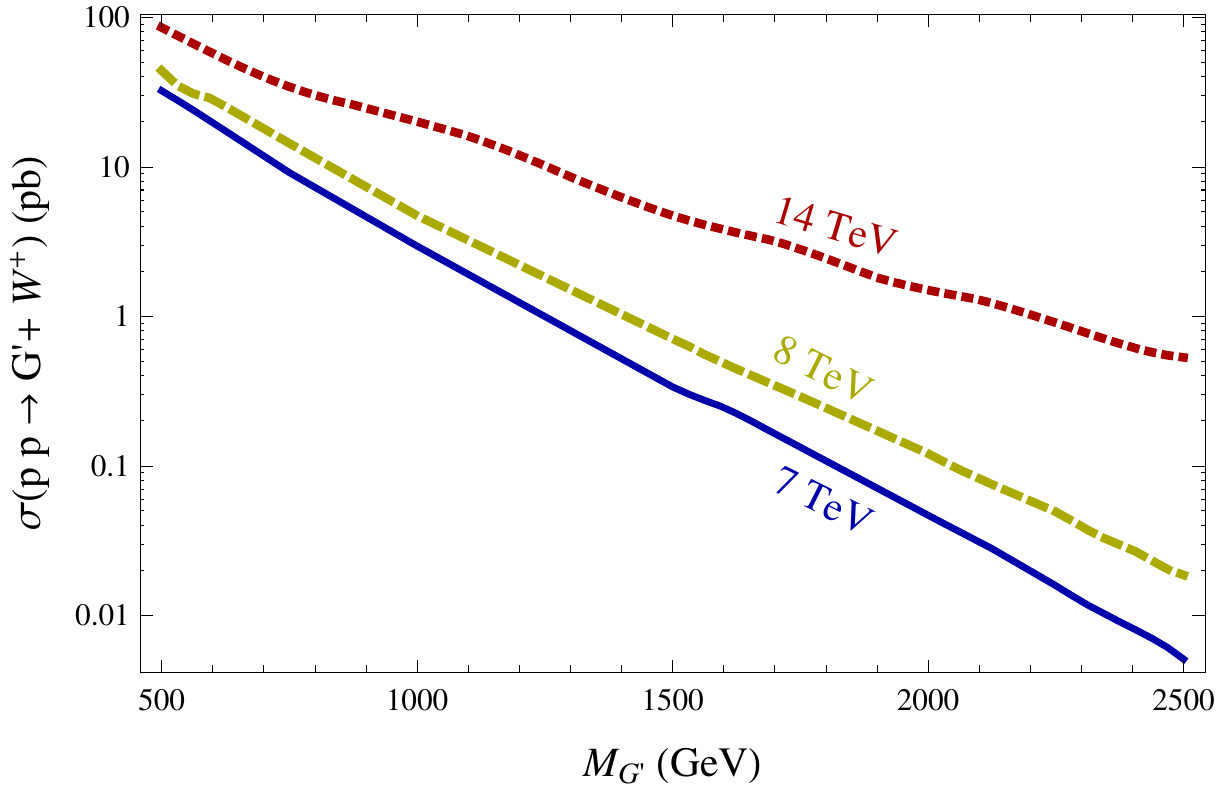} 
\caption{The production cross section of $G^\prime$ in association with $W^+$ at the LHC for three different center of mass energies. The couplings of $G^\prime$ to light quarks are chosen to be $|g^q_V|^2 +|g^q_A|^2 =2$.
}
\label{fig:GprimeW}
\end{center}
\end{figure}

Another signature for axigluons at the LHC is to study $G^\prime$ production together with a weak gauge boson. For the production of $p p \rightarrow G^\prime + W^{\pm}$, the final state is $2\,j + \ell + \missET$, which has a smaller background than the dijet resonance channel and thus may have a better discovery reach at the LHC. The partonic production cross section in leading order QCD is calculated to be~\cite{Bai:2010dj}
\be
\sigma(u \, \overline{d}\to G_\mu^\prime \,  W^+) =
\frac{8\pi\alpha\alpha_s\, (|g_V^q|^2 +  |g_A^q|^2)}{9\hat{s}^2\sin^2\!\theta_W}
\left[
\frac{\hat{s}^2 \!+\! (M^2_{G^\prime} \! + \! M^2_W)^2 \! }{\hat{s} - M^2_{G^\prime} - M^2_W} 
\ln\! \left( \frac{\hat{s}\!  - \! M^2_{G^\prime} \! -\!  M^2_W \! + \! M_\beta^2  }{2 M_{G^\prime} M_W }  \right) \! - \! M_\beta^2 \right] ~,
\ee
where 
\be
M_\beta^2 = \left[ \left(\hat{s} - M^2_{G^\prime} - M^2_W\right)^2 - 4 M^2_{G^\prime} M^2_W \right]^{1/2} ~.
\ee
Convoluting this partonic cross section with the MSTW \cite{Martin:2009iq} PDFs, we have the production cross sections at the LHC shown in Fig.~\ref{fig:GprimeW}.
The production cross section of $G^\prime + W^-$ is smaller than $G^\prime + W^+$ and is not shown in this figure. The major background comes from  $W^{+}$ plus jets from QCD and the discovery limit requires a detailed analysis, which is beyond the scope of this paper.

\subsection{Bottom Quark Forward-Backward Asymmetry}
\label{sec:bottomeFB}

An anomalously large forward-backward asymmetry in $t \bar t$ suggests the possibility of a similar asymmetry in $b \bar b$.  Models that preserve a custodial $SU(2)_R$ symmetry give rise to asymmetries of roughly equal magnitude in $t \bar t$ and $b \bar b$, so a measurement of the forward-backward asymmetry in $b$ quarks would provide further insight into the nature of possible new physics contributions~\cite{Sehgal:1987wi}.

Let us explore the feasibility of a $b \bar b$ asymmetry measurement using the top quark data as a guide.  Since the $t \bar t$ asymmetry was only significant for $M_{t \bar t} > 450$ GeV and $\Delta y > 1$, it is natural to assume that an asymmetry in $b$ quarks will be most visible in an analogous region.  Furthermore, in order to tag the sign of a $b$ quark, the $B$ meson must decay to a visible electron or a muon.  At CDF this requires $|y| < 1$ for any tagged $b$-jet, and in any case $b$-tagging itself quickly loses efficiency outside of this central region \cite{Acosta:2005zd}~\footnote{The $b$-tagging efficiency at D0 gradually decreases for $|y| > 1$ as can be seen from Fig. 28 of Ref.~\cite{Abazov:2010ab}. Therefore, D0 may have a better sensitivity for the $b$-quark forward backward asymmetry.}.   To estimate the number of events available for a $b \bar b$ asymmetry measurement with these rapidity and invariant mass constraints, we integrated MSTW PDFs \cite{Martin:2009iq} using the leading order QCD cross section, resulting in a cross section times acceptance of $0.3$ pb for $M_{b\bar b} > 450$~GeV.  So with a full data set of $10$ fb$^{-1}$, there would be about $3000$ $b \bar b$ events to work with.  However, it is possible that the $t \bar t$ asymmetry was dominated by the region $\Delta y > 1$ in part because for smaller $\Delta y$ the top decay makes it difficult to resolve the initial direction of the $t \bar t$ system.  Since we would not have this difficulty with the $b \bar b$ system, we might be less conservative and only assume $|y| < 1$, in which case the cross section times acceptance rises to $1.0$ pb, giving about $10^4$ $b \bar b$ events.

Now we face more difficult questions -- how accurately can we tag and measure the sign of one or both $b$ quarks, and what is the optimal procedure?  A dedicated analysis will be necessary to answer these questions, so here we will limit ourselves to pointing out some of the issues and making some very preliminary estimates.  

In an ideal world, it would be sufficient to simply find and measure the sign of a lepton inside of $b$-jets in dijet events.  In the real world, this is complicated by the probability of semi-leptonic $b$ decays, the efficiency and mis-tagging rate of $b$-tagging, $B^0$-$\bar B^0$ mixing, and by the possibility of leptons from charm decays inside of $b$-jets (resulting in mis-measurements of the sign of the $b$ quark).  These confounding factors might or might not make it worthwhile to require a $b$-tag for both jets, or perhaps to attempt to measure the sign of both jets in order to veto like-sign events.  Let us consider these effects in turn.  The probability of a semi-leptonic (electron or muon) $b$-decay is about $0.2$, and presumably this will be correlated with the $b$-tagging efficiency.  The $b$-tagging efficiency at CDF peaks at $p_T \approx 60$ GeV and then slowly drops off at high $p_T$, falling to about $0.3$ at $200$ GeV~\cite{Neu:2006rs}.  A given $b$-quark will hadronize into a $B_d$ meson about $40\%$  of the time, and into a $B_s$ meson about $11\%$  of the time; other hadronization products do not mix with their anti-particles.   So in total, a given high-energy $b$-quark will have a probability of about $0.13$ to mix into a $\bar B$ meson before it decays, and this effect will dilute our sample.  Many analyses address the issue of separating the decays of $D$ and $K$ mesons from $B$ decays for the purpose of studying these mesons.  It should be possible to separate the leptons from $B$ decays from the leptons from charm decays for the purpose $b$-quark signing at high energies \cite{Sehgal:1987wi} using a cut on the lepton's momentum transverse to the jet (see Fig. 22 of Ref.~\cite{Abazov:2010ab} for the muon $p_T$ distributions from $b$-quark and $c$-quark).  Several studies, such as \cite{Abazov:2006qp, Aaltonen:2009hz}, address the separation of charm and bottom mesons.

In the most optimistic scenario, it would be feasible to require only one $b$-tag.  Then one would look for a lepton in either jet and tag its sign; the most optimistic probability for finding a lepton would be $\sim 30\%$. The presence of this lepton should help to further reject non-$b$ dijet events.  If we make the conservative assumption that $b$-tagging is independent of lepton-finding, then the overall acceptance of the tagging procedure will be about $10\%$ .  With the requirement that $M_{b \bar b} > 450$ GeV, $|y|<1$ for both jets, and $\Delta y > 1$ we would be working with about $300$ events.  If these events have the same parton-level asymmetry as was observed for top quarks, namely $A^{t \bar t}_{FB} \approx 0.45$, then after convolving with the inevitable $B^0$-$\bar B^0$ mixing, we could expect about $200$ events with a positive asymmetry and $100$ events with a negative asymmetry.  Relaxing the $\Delta y$ requirement would increase the total number of events by a factor of $3$, but it would dilute the asymmetry by an unknown amount.  The largest and perhaps the most challenging unknown is the probability of fake wrong-sign leptons from charm decays inside of $b$-jets.  Even if a study with $M_{b \bar b} > 450$ GeV proves very challenging, it is natural to hope that a study of lower energy $b \bar b$ events will be feasible, because the total cross section would be far larger and the $b$-tagging efficiency somewhat greater.

Requiring two $b$-tags would increase the purity of the sample, but it might be even more useful to require two leptons.  At best this procedure accepts only $4\%$ of $b \bar b$ events, but it makes it possible to measure the sign of both $b$-jets and reject like-sign events, reducing the impurities from leptons from charm decays and from $B^0$-$\bar B^0$ mixing.  In the $t \bar t$ study it was possible to consider events with two $b$ tags as a cross check, so it is not unreasonable to think that a similar study will be possible in $b \bar b$.  

A detailed experimental analysis will be necessary to settle these issues, but we are optimistic that CDF and D0 could measure the forward-backward asymmetry in $b \bar b$.  Such a study would give us valuable insight into the nature of the $t \bar t$ asymmetry, and a positive result would greatly bolster the case for new physics.


\section{Discussion and conclusions}
\label{sec:conclusions}

In this paper we have performed a general study of color octet vector boson exchange as a possible source of the large top quark forward-backward asymmetry, 
$A_{FB}^{t\bar t}$, observed by CDF at high $t\bar t$ invariant masses. To set the tone for our subsequent discussions we first considered an effective field theory
with only one new field, $G^\prime$, and examined the couplings that such a field would need in order to explain $A_{FB}^{t \bar t}$  while avoiding other experimental 
constraints arising from ($i$) the $t\bar t$ total cross section as well as $t\bar t$ resonance searches 
at large $M_{t\bar t}$ at the Tevatron and, correspondingly, ($ii$) dijet resonance and contact interaction searches at the LHC. Given the large 
parameter freedom, we found that for a suitable choice of $G^\prime$ couplings all of these requirements can be simultaneously satisfied and 
would most likely lead to a rather wide state ($\Gamma/M >0.2$) with a mass in excess of $\simeq 1.5$ TeV. A state with such properties would very likely 
be found by various LHC searches performed during the coming year. 

Problems arise, however, when we try to construct a detailed model. 
An obvious first choice would be a two-site coset model based on $SU(3)_1\times SU(3)_2 / SU(3)_c$. This scenario has only two free parameters, and while it was found to produce a large value for 
$A_{FB}^{t\bar t}$ with the correct sign, the values required for the light quark couplings were necessarily large enough to produce a significant 
signal for dijet contact interactions at the LHC, beyond that allowed by current CMS constraints. This model is, then, highly disfavored.  We considered three-site generalizations, but they did not fare much better. 

Both the two and three site models necessarily coupled the axigluon(s) with equal strength to the top and light quarks.  Since we require a large coupling to the top quark to produce a large $A_{FB}^{t\bar t}$, these models necessarily led to large dijet contact interactions. 
Clearly, one needs to construct a model where the axigluon couplings to the top-quark can be large while maintaining small couplings to the light quarks;
one way to do this was to return to the original two-site model and introduce a new $Q=2/3$ vector-like fermion which mixes with the RH $u$-quark, introducing an additional free parameter. In this case we found that there were significant regions of parameter space that satisfied all of the experimental requirements with a rather light $G^\prime$ mass $\sim$ 1 TeV.  The LHC is capable of probing all of this 
parameter space during the coming year. Furthermore, if the new vector-like fermion has a mass of $1$- $2$ TeV, its pair-production and subsequent decay to $u+G^{\prime(*)} \rightarrow u + t\bar t, u + u\bar u$, with the $G^\prime$ being either on- or off-shell, could be visible at the LHC. 

Since in all of these scenarios the third generation fermion couplings to the new color octet vector states differed from those of the first two, FCNCs could 
become problematic when one rotates to the mass eigenstate basis. We derived the general form for the constraints that arise from mixing in the Kaon, $D$ and $B_q$ meson sectors. 
These are made somewhat more complicated by the generic existence of both left- and right-handed FC interactions, necessitating a full operator analysis. However, it is impossible to numerically evaluate these FCNC contributions in a completely model-independent 
fashion since there are, in general, four distinct unitary matrices which are necessary to rotate the $Q=2/3$ and $-1/3$ fermions from the weak to the mass 
eigenstate basis and only the experimental values of the CKM matrix (for LH charged currents) act as direct constraints. To perform this evaluation in a 
unique way requires that any model under consideration also provides a theory of flavor (which ours do not). If we assume that the left-handed and right-handed mixing matrices are identical, we found that all the $A^{t\bar t}_{FB}$ preferred regions for the models considered here are excluded. However, if the two right-handed mixing matrices are trivial, many axigluon models survive the flavor constraints. 

Since $t_L$ and $b_L$ lie in a single doublet and any model with an $SU(2)_R$ custodial symmetry relates $t_R$ and $b_R$ couplings, it is natural to 
ask if the asymmetry observed in the $t\bar t$ system might also manifest itself in the $b\bar b$ final state at large invariant masses at the Tevatron or 
the LHC. Given the level of statistics and the issues associated with $b$-tagging and signing plus possible dilutions arising from mixing and charm contamination 
it is not possible for us to easily evaluate how feasible such measurements might be. We encourage the collider collaborations to consider this possibility, 
especially if the large values of $A_{FB}^{t\bar t}$ are confirmed by future measurements. Clearly this would provide us with valuable information on the 
physics behind this asymmetry. 

We have constructed models of color octet vector bosons that explain the large value of $A_{FB}^{t \bar t}$ while remaining consistent with collider and flavor constraints, but it should be clear from the discussion above that they are not particularly elegant.  All of these models necessarily entail dijet or $t \bar t$ signatures that will be testable in the coming year at the LHC.  An immediate experimental prospect is worth a thousand models.

\subsection*{Acknowledgments} 
We would like to thank Dan Amidei, Brendan Casey, Tim Tait, and Jay Wacker for useful discussions and comments. SLAC is operated by Stanford University for the US Department of Energy under contract DE-AC02-76SF00515.

\providecommand{\href}[2]{#2}\begingroup\raggedright\endgroup

\end{document}